\documentstyle [12pt,eqsecnum,aps,amsfonts] {revtex}
\input epsf
\newcommand{\beq}{\begin{equation}}
\newcommand{\eeq}{\end{equation}}
\newcommand{\beqs}{\begin{eqnarray}}
\newcommand{\eeqs}{\end{eqnarray}}

\begin{document}
\tighten
\draft

\baselineskip 6.0mm

\title{Chromatic Polynomials and their Zeros and Asymptotic Limits for Families
of Graphs} 

\author{Robert Shrock\thanks{email: robert.shrock@sunysb.edu. invited talk 
at the 1999 British Combinatorial Conference BCC-99, Univ. of Kent, July 12, 
1999}}

\address{
Institute for Theoretical Physics  \\
State University of New York       \\
Stony Brook, N. Y. 11794-3840  \\
USA} 

\maketitle

\vspace{10mm}

\begin{abstract}

Let $P(G,q)$ be the chromatic polynomial for coloring the $n$-vertex graph $G$
with $q$ colors, and define $W=\lim_{n \to \infty}P(G,q)^{1/n}$.  Besides their
mathematical interest, these functions are important in statistical physics.
We give a comparative discussion of exact calculations of $P$ and $W$ for a
variety of recursive families of graphs, including strips of regular lattices
with various boundary conditions and homeomorphic expansions thereof.
Generalizing to $q \in {\mathbb C}$, we determine the accumulation sets of the
chromatic zeros constituting the continuous loci of points on which $W$ is
nonanalytic.  Various families of graphs with the property that the chromatic
zeros and/or their accumulation sets (i) include support for $Re(q) < 0$; (ii)
bound regions and pass through $q=0$; and (iii) are noncompact are discussed,
and the role of boundary conditions is analyzed. Some corresponding results are
presented for Potts model partition functions for nonzero temperature,
equivalent to the full Tutte polynomials for various families of graphs. 

\end{abstract}

\vspace{16mm}

\pagestyle{empty}
\newpage

\pagestyle{plain}
\pagenumbering{arabic}
\renewcommand{\thefootnote}{\arabic{footnote}}
\setcounter{footnote}{0}

\section{Introduction}

We consider connected  
graphs $G$ without loops or multiple bonds and denote the number of
vertices as $n=v(G)$, the edges as $e(G)$, the girth as $g$, the number of
circuits with this minimal length as $k_g$, and the chromatic number as
$\chi(G)$.  The chromatic polynomial $P(G,q)$ counts the number of ways that
one can color the graph $G$ with $q$ colors such that no two adjacent vertices
have the same color \cite{birk} (for reviews, see \cite{rrev}-\cite{bbook}).
The minimum number of colors needed for this coloring, i.e., the smallest
positive integer $q$ for which $P(G,q)$ is nonzero, is the chromatic number, 
$\chi(G)$.  Besides its intrinsic mathematical interest, the chromatic
polynomial has an important connection with statistical mechanics since it is
the zero-temperature partition function of the $q$-state Potts antiferromagnet
(AF) \cite{potts,wurev} on $G$: 
\beq 
P(G,q)=Z(G,q,T=0)_{PAF} 
\label{pz}
\eeq

We shall consider recursively defined families of graphs here, i.e., roughly
speaking, those that can be formed by successive additions of a given
subgraph. A precise definition of a recursive family $G_m$ depending on a
positive integer parameter $m$ is a sequence of graphs for which the chromatic
polynomials $P(G_m,q)$ are related by a linear homogeneous recurrence relation
in which the coefficients are polynomials in $q$ \cite{bbook,bds}. These
include strips of regular lattices with various boundary conditions, chains of
polygons linked in various ways, joins of such graphs with another graph such
as a complete graph $K_p$, and families obtained from these by modifications
such as removal of some edges or additions of degree-two vertices, i.e.,
homeomorphic expansions. These families of graphs may depend on several
parameters (e.g., width, length, number of homeomorphic expansions, etc.), and
there can be several ways in which one can obtain the limit $n \to \infty$.
Let us concentrate on one such parameter, such as the length of a strip of a
regular lattice of fixed width or polygon chain, $m$, so that $n$ is a linear
function of $m$. 

Just as the chromatic polynomial counts the total number of ways of coloring
the graph $G$ with $q$ colors subject to the above constraint, so also it is 
useful to define a quantity that measures the number of ways of performing this
coloring per site, in the limit where the number of vertices goes to infinity. 
Denoting the formal limit $\{G\}=\lim_{n \to \infty}G$, we 
have\footnote{\tighten{\footnotesize{ At certain special points $q_s$
(typically $q_s=0,1,.., \chi(G)$), one has the noncommutativity of limits
$\lim_{q \to q_s} \lim_{n \to \infty} P(G,q)^{1/n} \ne \lim_{n \to \infty}
\lim_{q \to q_s}P(G,q)^{1/n}$, and hence it is necessary to specify the order
of the limits in the definition of $W(\{G\},q_s)$ \cite{w}.  We use the first
order of limits here; this has the advantage of removing certain isolated
limits here; this has the advantage of removing certain isolated
discontinuities in $W$.}}}
\beq
W(\{G\},q)=\lim_{n \to \infty} P(G,q)^{1/n}
\label{w}
\eeq 
With the order of limits specified in the footnote, this limit exists for the 
recursive families of graphs considered here, as will be discussed further
below. The function $W$ is the ground state degeneracy per vertex in the $n \to
\infty$ limit.  The quantity $S_0=k_B \ln W$, where $k_B$ is the Boltzmann
constant, is defined as the ground state entropy.  The Potts antiferromagnet
has the interesting feature that it exhibits nonzero ground-state entropy 
$S_0 \ne 0$ (without frustration) for sufficiently large $q$ on a given 
lattice or graph and is thus an exception to the third law of thermodynamics
\cite{al}.  
This is equivalent to a ground state degeneracy per site $W > 1$.  For example,
for the square lattice, $W=(4/3)^{3/2}$ \cite{lieb}. 
We recall that, with $n=v(G)$, a general form for the chromatic polynomial of a
connected graph $G$ is 
\beq
P(G,q) = \sum_{j=0}^{n-1}(-1)^j h_{n-j}q^{n-j}
\label{p}
\eeq
where $h_{n-j}$ are positive integers, with \cite{rtrev,m}
$h_{n-j}={e \choose j}$ for $0 \le j < g-1$ (whence $h_n=1$ and $h_{n-1}=e$)
and $h_{n-(g-1)}={e \choose g-1}-k_g$. 

Although in the Hamiltonian formulation of the $q$-state Potts model or the
mathematical context of coloring a graph with $q$ colors, $q$ must be a
non-negative integer, once one has the function $P(G,q)$, one can generalize
the quantity $q$ from ${\mathbb Z}_+$ to ${\mathbb C}$.  A subset of the zeros
of $P$ in the $q$ plane (chromatic zeros) may form a continuous accumulation
set in the $n \to \infty$ limit, denoted ${\cal B}$, which is the continuous
locus of points where $W(\{G\},q)$ is nonanalytic (${\cal
B}$ may be null, and $W$ may also be nonanalytic at certain discrete points).
The maximal region in the complex $q$ plane to which one can analytically
continue the function $W(\{G\},q)$ from the range of positive integer values
where there is nonzero ground state entropy is denoted $R_1$.  The maximum
value of $q$ where ${\cal B}$ intersects the (positive) real axis is labelled
$q_c(\{G\})$.  Thus, region $R_1$ includes the positive real axis for $q >
q_c(\{G\})$.  We have calculated ${\cal B}$ for a variety of families of graphs
and shall present rigorous results, some observations, and some conjectures
here.  Part of the interest in this area of research is that it combines, in a
fruitful way, both theoretical physics and three areas of mathematics: graph
theory, complex analysis, and algebraic geometry. In addition to the works
already cited, some previous relevant papers are \cite{whit}-\cite{tk}. 

The recursive families of graphs considered here 
include strips of regular lattices with various boundary conditions, chains of
polygons linked in various ways, joins of such graphs with another graph such
as a complete graph $K_p$, and families obtained from these by modifications
such as removal of some edges or additions of degree-two vertices, i.e., 
homeomorphic expansions.  
These families of graphs may depend on several parameters, and there can be
several ways in which one can obtain the limit $n \to \infty$.  Concentrating
on one such parameter, such as the length of a strip of a regular lattice or
polygon chain, $m$, and denoting the graph as $(G_s)_m$, a general form for 
$P((G_s)_m,q)$ is
\beq
P((G_s)_m,q) = \sum_{j=1}^{N_a} c_j(q)a_j(q)^m
\label{pgsum}
\eeq 
where $c_j(q)$ and the $N_a$ terms $a_j(q)$ (which we shall also denote
equivalently as $\lambda_j(q)$) depend on the type of strip graph $G_s$ but are
independent of $m$.  We define a term $a_\ell$ as ``leading'' ($\ell$) if it
dominates the $n \to \infty$ limit of $P(G,q)$. The locus ${\cal B}$ occurs
where, as one changes $q$, there is an abrupt, nonanalytic change in $W$ as the
leading terms $a_\ell$ in eq. (\ref{pgsum}) changes.  Hence, ${\cal B}$ is the
solution to the equation of degeneracy of magnitudes of leading terms,
$|a_\ell|=|a_{\ell^\prime}|$.  It follows that $|W|$ is finite and continuous,
although nonanalytic, across ${\cal B}$.  In general, for the families of
graphs considered here, ${\cal B}$ is an algebraic curve, since it is the locus
of solutions to the equation $|a_\ell|=|a_{\ell^\prime}|$ and the terms
$a_\ell$, $a_{\ell^\prime}$ are algebraic functions of $q$. From (\ref{pgsum}),
with the ordering as given in the footnote, one can show constructively that 
the limit (\ref{w}) exists; it is
given by $W(\{ G_s\},q)=(a_\ell)^t$ for $q \in R_1$, and $|W(\{
G_s\},q)|=|a_\ell|^t$ for $q \not\in R_1$, where $\lim_{m \to \infty} m/n = t$
and $a_\ell$ is the term among the $a_j$, $j=1,...,N_a$ with maximum magnitude
at the given value of $q$. Thus, a number of
interesting questions can be asked and answered about its properties.

Concerning the coefficients $c_j$, we note that
\beq
c_{\ell}(q) = 1 \quad {\rm if} \quad a_{\ell}(q) \quad {\rm is \ dominant \
in} \ \ R_1
\label{cr1}
\eeq
This is proved by using the fact that region $R_1$ is defined to include the
region of large real $q$, and then requiring the expression (\ref{pgsum}) to
match the general expression (\ref{p}) and satisfy the property that the
coefficient of the leading term, $q^n$, is unity. 
For some families of graphs, the $c_j$ and
$a_j$ are polynomials in $q$.  However, there are also families of graphs,
such as the $L_y=3$ cyclic and M\"obius strips of the square and kagom\'e
lattices \cite{wcy,pm} (see below for notation)
which have nonpolynomial algebraic $a_j$'s with
polynomial $c_j$'s, and families of graphs, such as certain strip graphs of
regular lattices with free transverse and longitudinal boundary conditions
\cite{strip,hs}, as well as 
the $L_y=2$ M\"obius strips of the triangular lattice and asymmetric
homeomorphic expansions of the $L_y=2$ M\"obius square strip \cite{wcy,pm} for 
which some $a_j$'s and their respective $c_j$'s are both algebraic, 
nonpolynomial functions of $q$.  When the $a_j$'s are algebraic nonpolynomial
functions of $q$, this happens because these are roots of an algebraic 
equation of degree 2 or higher.  There are then two 
possibilities: first, the roots may enter in a symmetric manner, 
and, because of a theorem on symmetric polynomial functions discussed below,
their coefficients are all equal and are polynomial functions of $q$.  In
particular, if one of these $a_j$'s is leading in $R_1$, then the
coefficient functions for all of the roots are equal to unity, in accordance
with (\ref{cr1}).  This happens, for example, for the $L_y=3$ cyclic and
M\"obius strips of the square lattice, where the leading term, 
$a_{sq,6} \equiv \lambda_{sq,6}$ in eq. (\ref{psqly3}) is
the root of a quadratic equation (see eq. (\ref{lambda67})) and enters the
chromatic polynomial together with the other root of this equation, each with
the same coefficient, which is unity.  A second possibility is that 
the nonpolynomial algebraic roots enter in (\ref{pgsum}) in a manner that is 
not a symmetric
function of these roots; in these cases, the corresponding coefficient
functions are also nonpolynomial algebraic functions of $q$.  For example, in
many strips with free transverse and longitudinal boundary conditions
\cite{strip,hs} and in the 
the M\"obius strip of the triangular lattice and the homeomorphic expansion 
of the M\"obius strip of the square lattice, one finds \cite{wcy} that two
of the $a_j$'s are roots of a quadratic equation of the form 
$a_{\pm} = R \pm \sqrt{S}$, and they occur in the chromatic polynomial 
in the form $(a_{+})^m - (a_{-})^m$; for these families, this
combination is multiplied by the factor $1/(a_{+}-a_{-})$, i.e.,
the coefficient functions contain the respective factors $\pm 1/\sqrt{S}$. 
In all cases, of course, these combine to yield a $P(G,q)$ that is a polynomial
in $q$.  

Two other general result that follow from eqs. (\ref{w}), (\ref{p}), and 
(\ref{pgsum}) are as follows.  As $m,n \to \infty$, let $r=n/m$, which depends
on $G_s$.  Then
\beq
a_\ell \sim q^r \quad {\rm for} \quad q \in R_1 \ , \ \ |q| \to \infty
\label{aell}
\eeq
\beq
W = (a_\ell)^{\frac{1}{r}} \quad {\rm for} \quad q \in R_1 
\label{wt}
\eeq
so that $\lim_{|q| \to \infty; \ q \in R_1} q^{-1}W = 1$.

\section{Calculation of Chromatic Polynomials}

For a recursively defined family of graph $G_s$ our calculational method is to
use the deletion-contraction theorem iteratively.   For some families of graphs
we have directly solved for the chromatic polynomials from the recursion
relations that follow from this iterative procedure. A related method that 
that we have found to be useful employs a generating function and is described
below \cite{strip,hs}.  Let us consider strips of various
lattices with arbitrary length $L_x=m$ vertices and fixed width $L_y$ vertices
(with the longitudinal and transverse directions taken to be $\hat x$ and $\hat
y$). The chromatic polynomials for the cyclic and M\"obius strip graphs of the
square lattice were calculated for $L_y=2$ in \cite{bds} by means of iterative
deletion-contraction operations, and subsequently via a transfer matrix method
in \cite{bm} and a coloring matrix method in \cite{b77} (see also 
\cite{matmeth}).  We have extended this to the corresponding cyclic \cite{wcy}
and M\"obius \cite{pm} $L_y=3$ strips.  For each graph family, once one has
calculated the chromatic polynomial for arbitrary length, one can take the
limit of infinite length and determine the accumulation set ${\cal B}$. 
After studies of the chromatic zeros for $L_y=2$ in
\cite{bds,readcarib,read91}, $W$ and ${\cal B}$ were determined for this case
in \cite{w} and for $L_y=3$ in \cite{wcy,pm}. An interesting question concerns
the effect of boundary conditions (BC's), and hence graph topology, on $P$,
$W$, and ${\cal B}$. We use the symbols FBC$_y$ and PBC$_y$ for free and
periodic transverse boundary conditions and FBC$_x$, PBC$_x$, and TPBC$_x$ for
free, periodic, and twisted periodic longitudinal boundary conditions.  The
term ``twisted'' means that the longitudinal ends of the strip are identified
with reversed orientation. These strip graphs can be embedded on surfaces with
the following topologies: (i) (FBC$_y$,FBC$_x$): strip;
(ii) (PBC$_y$,FBC$_x$): cylindrical; (iii) (FBC$_y$,PBC$_x$): cylindrical
(denoted cyclic here); (iv) (FBC$_y$,TPBC$_x$): M\"obius; (v)
(PBC$_y$,PBC$_x$): torus; and (vi) (PBC$_y$,TPBC$_x$): Klein bottle.\footnote{
\footnotesize{These BC's can all be implemented in a manner that is uniform in
the length $L_x$; the case (vii) (TPBC$_y$,TPBC$_x$) with the topology of the
projective plane requires different identifications as $L_x$ varies and is 
not considered here.}} 
We have calculated $P$, $W$, and ${\cal B}$ for a variety of 
strip graphs having BC's of type (i) \cite{strip}-\cite{w2d}, (ii)
\cite{w2d}, (iii) \cite{pg}-\cite{nec}, (iv) \cite{pg,pm}.
Recently, strip graphs of the square lattice with 
BC's of torus (v) and Klein bottle (vi) type have also been studied \cite{tk}. 
In addition to strips of regular lattices, we have calculated $P$, $W$, and
${\cal B}$ for cyclic graphs of polygons linked in various manners \cite{nec}
and homeomorphic expansions of strip graphs \cite{hs,pg,wcy}. 

We proceed to describe our generating function method for calculating chromatic
polynomials.  The generating function is denoted. $\Gamma(G_s,q,x)$, and 
the chromatic polynomials for the strip of length $L_x=m$
are determined as the coefficients in a Taylor series expansion of this
generating function in an auxiliary variable $x$ about $x=0$:
\beq
\Gamma(G_s,q,x) = \sum_{m=m_0}^{\infty}P((G_s)_m,q)x^{m-m_0}
\label{gamma}
\eeq
where $m_0$ depends on the type of strip graph $G_s$ and is naturally chosen as
the minimal value of $m$ for which the graph is well defined.  
The generating functions $\Gamma(G_s,q,x)$ are rational functions of the form 
\cite{strip} 
\beq
\Gamma(G_s,q,x) = \frac{{\cal N}(G_s,q,x)}{{\cal D}(G_s,q,x)}
\label{gammagen}
\eeq
with
\beq
{\cal N}(G_s,q,x) = \sum_{j=0}^{d_{\cal N}} A_{G_s,j}(q) x^j
\label{n}
\eeq
and
\beq
{\cal D}(G_s,q,x) = 1 + \sum_{j=1}^{d_{\cal D}} b_{G_s,j}(q) x^j
\label{d}
\eeq
where the $A_{G_s,i}$ and $b_{G_s,i}$ are polynomials in $q$ (with no common
factors).  
Writing the denominator of $\Gamma(G_s,q,x)$ in factorized form, we have
\beq
{\cal D}(G_s,q,x) = \prod_{j=1}^{deg_x{\cal D}}(1-\lambda_{G_s,j}(q)x)
\label{lambdaform}
\eeq
One can then calculate $c_j$ and $a_j$ in (\ref{pgsum}) in terms of these
quantities:
\beq
P(G_m,q) = \sum_{j=1}^{d_{\cal D}} \Biggl [ \sum_{s=0}^{d_{\cal N}}
A_s \lambda_j^{d_{\cal D}-s-1} \Biggr ]
\Biggl [ \prod_{1 \le i \le d_{\cal D}; i \ne j}
\frac{1}{(\lambda_j-\lambda_i)} \Biggr ] \lambda_j^m
\label{chrompgsumlam}
\eeq
We have proved that for strip graphs with (FBC$_x$,FBC$_y$) boundary
conditions, all of the $\lambda_j$'s contribute to $P$, i.e. none of the
coefficients in (\ref{chrompgsumlam}) vanish \cite{strip,hs}.  However, for
some graphs with  (FBC$_x$,PBC$_y$) or (FBC$_x$,TPBC$_y$) boundary conditions,
we have also proved that certain coefficients $c_j$ in eq. 
(\ref{chrompgsumlam}) do vanish, so that the corresponding $\lambda_j$;s do 
not contribute to $P$ and $N_a \le deg_x({\cal D})$ \cite{wcy,pm}. In general, for the $\lambda_j$'s that do contribute,
\beq
a_j = \lambda_j
\label{alam}
\eeq

We show here how the nonpolynomial algebraic roots in the various $P$'s yield
polynomials in $q$.  As is evident from (\ref{lambdaform}), for a given strip,
the $\lambda_j$'s arise as roots of the equation ${\cal D}=0$.  In general,
${\cal D}$ contains some number of factors of linear, quadratic, cubic, etc.
order in $x$.  Consider a generic factor in ${\cal D}$ of $r$'th degree in $x$:
\ $(1+f_1x+f_2x^2+...+f_rx^r)$, where the $f_j$'s are polynomials in $x$.  This
yields $r$ \ $\lambda_\ell$'s as roots of the equation
$\xi^r+f_1\xi^{r-1}+...+f_r=0$. The expressions in $P$ involving these roots
are symmetric polynomial functions of the roots, namely terms of the form \beq
s_m = \sum_{\ell=1}^r (\lambda_\ell)^m
\label{s}
\eeq
and, for the M\"obius strips of the $L_y=2$ triangular lattice and asymmetric 
homeomorphic expansions of the $L_y=2$ M\"obius square strip, as well as 
strips with (FBC$_y$,FBC$_x$), terms of the form
\beq
\frac{(\lambda_+)^m-(\lambda_-)^m}{\lambda_+ -\lambda_-} =
\sum_{k=0}^{m-1}(\lambda_+)^{m-1-k}(\lambda_-)^k \ .
\label{ldif}
\eeq 
We can then apply a standard theorem (e.g. \cite{uspensky}) which states
that a symmetric polynomial function of the roots of an algebraic equation is a
polynomial in the coefficients, here $f_\ell$, $\ell=1,..,r$; hence this
function is a polynomial in $q$.  For example, for $s_r$, one has the
well-known formulas (due to Newton) $s_1=-f_1$, $s_2=f_1^2-2f_2$,
$s_3=-f_1^3+3f_1f_2-3f_3$, etc.  It becomes progressively more and more
time-consuming to calculate these $s_r$'s for large $m$ and hence large $r$; it
is here that one uses the full power of the generating function method, which
immediately yields the chromatic polynomial without the necessity of having to
go through the intermediate stage of calculating the $\lambda_j$'s and then use
Newton identities to get rid of algebraic roots and obtain the final
polynomial.

\section{General Results on ${\cal B}$}

One can ask a number of questions about the properties of ${\cal B}$.  We take
the opportunity here to give a unified discussion of these which includes
results from our various studies together with some new remarks. 

\begin{enumerate}

\item

Does ${\cal B}$ have general symmetries?  The answer is yes; a basic symmetry
is that 
\beq
{\cal B}(q) = {\cal B}(q^*)
\label{bsym}
\eeq
i.e., ${\cal B}$ is invariant under complex conjugation in the $q$ plane.  This
follows from the fact that the chromatic zeros have this property, which, in 
turn, follows from the fact that the coefficients in the chromatic polynomial 
are real (cf. eq. (\ref{p})). 

\item 

What is the dimensionality of ${\cal B}$ in the $q$ plane?  As noted after
(\ref{pgsum}), one proves 
from the definition of ${\cal B}$ as the solution of the equality in magnitude
of leading terms $a_j$ in eq. (\ref{pgsum}), together with the fact that the
$a_j$'s are algebraic functions of $q$, that (in cases where it is not the
empty set) that ${\cal B}$ is an algebraic curve (including possible line
segments on the real $q$ axis), so 
\beq
dim({\cal B}) = 1
\label{dimb}
\eeq

\item

What is the topology of the curve ${\cal B}$ and the associated ``region
diagram'' in the $q$ plane?  Does ${\cal B}$ separate the $q$ plane into two or
more regions or not?   Does ${\cal B}$ cross the real $q$ axis, and, if so, at
which points?  From our studies \cite{strip,hs,pg,wcy,nec,pm}, we 
arrive at the following inference, which we state as a conjecture:  

{\it Conjecture}.\ \ For a family of graphs $G_s$ with well-defined lattice
structure\footnote{\footnotesize{By well-defined lattice structure we mean that
the vertices and edges of the graph can be put into a 1-1 correspondence with
the vertices and edges of a section of a regular lattice.}}, a sufficient
condition for ${\cal B}$ to separate the $q$ plane into different regions is
that $G_s$ contains at least one global circuit, defined as a route following a
lattice direction which has the topology of $S^1$ and a length $\ell_{g.c.}$
that goes to infinity as $n \to \infty$.  In the context of strip graphs, this
is equivalent to having PBC$_x$, i.e., periodic boundary conditions (or
TPBC$_x$, twisted periodic boundary conditions) in the direction in which the
strip length goes to infinity as $m \to \infty$.

The presence of a global circuit in a family of graphs is not a necessary
condition for ${\cal B}$ to enclose regions, as was shown in \cite{strip2}. We
concentrate here on lattice strips because of the connection to statistical
mechanics; however, we also have studied families of graphs which do not have a
lattice structure but, in the $n \to \infty$ limit, yield loci ${\cal B}$ that
separate the $q$ plane into regions.  These include cyclic chains of polygons
\cite{pg,nec} (for which the analogue of the global circuit is the route around
the chain) and certain families of graphs with noncompact ${\cal B}$
\cite{read91,wa,wa3,wa2,sokal}; these are discussed below.  A related
conjecture is that, for a family of graphs $G_s$ with well-defined lattice
structure, a necessary and sufficient condition for ${\cal B}$ to separate the
$q$ plane into regions and pass through $q=0$ is that $G_s$ contains at least
one global circuit.  Indeed, all of the families of graphs that we have studied
that contain global circuits also yield loci ${\cal B}$ that pass through $q=2$
as well as $q=0$.  In contrast, for strip graphs with (FBC$_y$,FBC$_x$) studied
in \cite{strip,strip2}, ${\cal B}$ consists of arcs that, in the simplest
cases, do not enclose regions (c.f. Figs. 3-9 of \cite{strip}, Figs. 2(a),3(a)
of \cite{strip2}, and in other cases do enclose regions (c.f. Figs. 2(b), 3(b)
of \cite{strip2}), but do not pass through $q=0$.  We find that as the width of
the strip increases, these arcs tend to elongate and move toward each other,
thereby suggesting that if one considered the sequence of strip graphs of this
type with width $L_y$ and length $L_x$ (having taken the limit $L_x \to \infty$
to obtain a locus ${\cal B}$ for each member of this sequence), then in the
limit $L_y \to \infty$, the arcs would close to form a ${\cal B}$ that passes
through $q=0$ in such a way as always to separate the $q$ plane into different
regions. The interesting feature of the cyclic and M\"obius strips and cyclic
polygon chain graphs \cite{pg,wcy,nec,pm} is that when the graphs contain a
global circuit, this property of ${\cal B}$ that it passes through $q=0$ in
such a manner as to separate the $q$ plane into regions already occurs for
finite $L_y$. This means that the $W$ functions of these graphs with cyclic and
twisted cyclic longitudinal boundary conditions already exhibit a feature which
is expected to occur for the ${\cal B}$ for the $W$ function of the full
two-dimensional lattice.  This expectation is supported by the calculation of
$W$ and ${\cal B}$ for the 2D triangular lattice with cylindrical boundary
conditions by Baxter \cite{baxter}.  Recently, in \cite{tk} with Biggs, we have
found the same property for strips of the square lattice with (PBC$_y$,PBC$_x$)
(torus) and (PBC$_y$,TPBC$_x$) (Klein bottle) boundary conditions.

\item 

Does ${\cal B}$ have singular points in the technical terminology of algebraic
geometry, such as (i) multiple points, where several branches of the curve
intersect without crossing or cross each other, or (ii) endpoints?  All of
these possibilities are realized.  There are cases where there are no multiple
points, such as for the circuit and $p$-wheel graphs (defined in eq.
(\ref{pwheel}) below) and some polygon chain graphs.  There are also cases
where there are (a) multiple points associated with intersecting but
non-crossing branches, such as the $L_y=2$ and $L_y=3$ cyclic and M\"obius
strips of the square lattice and the $L_y=2$ strips of the kagom\'e ($3 \cdot 6
\cdot 3 \cdot 6$) lattice \cite{w,wcy,pm}; (b) multiple points associated with
crossing curves, such as the $L_y=2$ cyclic and M\"obius strips of the
triangular lattices and certain families of cyclic polygon graphs; and (c)
multiple points associated with both branch intersections and crossings, such
for as homeomorphic expansions of cyclic square strips \cite{pg}. We have also
shown that homogeneous strips of regular lattices with free transverse and
longitudinal boundary conditions have loci ${\cal B}$ with endpoints
\cite{strip} (see further below).

\item 

Does ${\cal B}$ consist of a single connected component, or several distinct
components?  In cases where ${\cal B}$ does not contain any multiple
points, the number of regions, $N_{reg.}$
and the number of connected components on ${\cal B}$ satisfy the relation
$N_{reg.}=N_{comp.}+1$.  The Harnack theorem \cite{shaf} gives the upper bound 
$N_{comp.} \le h+1$, where $h$, the genus of the algebraic curve comprising 
${\cal B}$, is $h=(d-1)(d-2)/2$.  However, as we have discussed in
\cite{wa2,nec}, this is a very weak bound.  For example, for the cyclic 
polygon chain family denoted $(e_1,e_2,e_g)=(2,2,1)$ in \cite{nec}, we showed
that $N_{comp.}=2$, while the Harnack theorem gives the upper bound 
$N_{comp.} \le 37$.  

\item 

  Although ${\cal B}$ is the continuous accumulation set of a subset of the
chromatic zeros, these zeros do not, in general, lie precisely on this
asymptotic locus ${\cal B}$ for finite $n$.  Are there families of graphs for
which all or a subset of the zeros do lie exactly on ${\cal B}$ for finite $n$?
We have answered this in the affirmative and have constructed a family of
graphs, which we call $p$-wheels, and proved that except for a finite subset of
chromatic zeros at certain nonnegative integer values (see below), all of the 
chromatic zeros lie exactly on ${\cal B}$ \cite{wc}.  Recall that the join of
two graphs $G$ and $H$ is defined as the graph consisting of copies of $G$ and
$H$ with additional edges added joining each vertex of $G$ to each vertex of 
$H$ \cite{bbook}.  A $p$-wheel is the join 
\beq
(Wh)^{(p)}_n = K_p + C_{n-p}
\label{pwheel}
\eeq 
so that $(Wh)^{(0)}_n$ is the circuit graph, $C_n$, $(Wh)^{(1)}_n$ is the
usual wheel graph, etc.  The locus ${\cal B}$ is the circle
\beq
|q-p-1|=1
\label{pwheelb}
\eeq
and the real chromatic zeros occur at $q=0,1,...p+1$ for $n-p$ even and
$q=0,1,...,p+2$ for $n-p$ odd, while the complex zeros all lie on the above
unit circle.  

\item 

What is the density of the zeros on ${\cal B}$?  In special cases, in 
particular, the $p$-wheel graphs, it is a constant.  In general, it varies
as one moves along ${\cal B}$ for a given family of graphs.  Our calculations
of chromatic zeros answer this question for specific families. 

\item 

 For many years, no examples of chromatic zeros were found with negative real
parts, leading to the conjecture that $Re(q) \ge 0$ for any chromatic zero
\cite{farrell}.  Although Read and Royle later showed that this conjecture is
false \cite{read91}, very few cases of graphs with chromatic zeros having
$Re(q) < 0$ are known, and the investigation of such cases is thus valuable for
the insight it yields into properties of chromatic zeros.  Note that the
condition that a graph has some chromatic zeros with $Re(q) < 0$ is a necessary
but not sufficient condition that it has an accumulation set ${\cal B}$ with
support for $Re(q) < 0$.  We have proved that certain homeomorphic expansions
of square strip graphs with the boundary conditions (FBC$_y$,FBC$_x$) \cite{hs}
and (FBC$_y$,PBC$_x$) \cite{pg}, of length $m$ units, have chromatic zeros with
$Re(q) < 0$ for sufficiently large $m$ and that they have loci ${\cal B}$ with
support for $Re(q) < 0$.  Since the homeomorphically expanded lattice strips in
\cite{hs} do not have global circuits, this shows that these circuits are not a
necessary condition for a recursive family to have chromatic zeros and
${\cal B}$ with support for $Re(q) < 0$, as is also shown for lattice strips
with $(PBC_y,FBC_x)$ in \cite{baxter}, falsifying a conjecture in \cite{strip}.
We have also shown the existence of chromatic zeros and ${\cal B}$ with $Re(q)
< 0$ for the $L_y=3$ cyclic and M\"obius strips of the square lattice and the
$L_y=2$ cyclic and M\"obius strips of the kagom\'e $(3 \cdot 6 \cdot 3 \cdot
6)$ lattice \cite{wcy,pm}. A different example is families of graphs with
noncompact ${\cal B}$ \cite{w,wa,wa3,wa2}; with these, we constructed cases
where the chromatic zeros have arbitrarily large negative $Re(q)$.

\item 

Is ${\cal B}$ compact (which in our context is synonymous with the property of
boundedness) in the $q$ plane, or does it extend to complex infinity in some
directions?  This is related to the question of boundedness of the magnitudes
of chromatic zeros.  In \cite{read91} Read and Royle gave an example of a graph
with a noncompact ${\cal B}$, namely the bipyramid graph.  We have studied the
question of the compactness of ${\cal B}$ in a series of papers
\cite{w,wa,wa3,wa2} (see also \cite{sokal}).  A necessary and sufficient
condition such that ${\cal B}$ is compact in the $q$ plane is obtained as
follows: one simply reexpresses the degeneracy equation for leading $a_j(q)$'s
in terms of the variable $z=1/q$ and determines if this has a solution for
$z=0$.  By constructing and studying families of graphs with noncompact loci
${\cal B}$, we have elucidated the geometrical conditions leading to this
noncompactness.  Our studies show that a necessary condition is that some
vertex has a degree $\Delta$ which goes to infinity as $n \to \infty$. However,
this is obviously not a sufficient condition for noncompactness of ${\cal B}$,
as is shown, for example, by the $p$-wheel graphs $(Wh)^{(p)})_n$, in which the
$p$ vertices in the complete graph $K_p$ have degree $\Delta$ that goes to
infinity as $n \to \infty$ but for which ${\cal B}$ is compact
(eq. (\ref{pwheelb}).  From our studies we have found that in all cases of
families with noncompact ${\cal B}$, the graphs have the common feature that,
in the limit as $n \to \infty$, they contain an infinite number of different,
non-overlapping non-self-intersecting circuits, each of which passes through
two or more nonadjacent vertices.  We are led to propose this as a conjecture
for the condition on a graph family such that it has a noncompact ${\cal B}$
\cite{wa2,wa3}.

\item 

The above conjecture also leads to the following corollary: A sufficient (not
necessary) condition for a family of graphs to have a compact, bounded locus
${\cal B}$ is that it is a regular lattice \cite{wa,wa3,wa2}.  Clearly, if
${\cal B}$ is noncompact, passing through the origin of the $z$ plane, where
$z=1/q$, then the function $q^{-1}W$ has no large-$q$ expansion, i.e., no
Taylor series expansion in the variable $z$ around the point $z=0$.  Our
conjecture is in accord with the derivation of large-$q$ expansions for regular
lattices \cite{ser}.

\item

Another feature that we find is that for families of graphs that (a) contain
global circuits, (b) cannot be written as the join $G=K_p + H$, where $K_p$ is
the complete graph on $p$ vertices, and (c) have compact ${\cal B}$, this locus
passes through $q=0$ and crosses the positive real axis, thereby always
defining a $q_c$. Note that ${\cal B}(\{K_p + H\},q)={\cal B}(\{H\},q-p)$,
i.e. the ${\cal B}$ for $\{K_p + H\}$ is the same as that for $\{H\}$ shifted a
distance $p$ to the right in the $q$ plane.  Hence, examples of families of
graphs with loci ${\cal B}$ that do not pass through $q=0$ include these joins,
where $H$ is a family whose ${\cal B}$ does pass through zero. Other examples
of graph families with ${\cal B}$ not including $q=0$ are those with noncompact
${\cal B}$ \cite{wa,wa3,wa2}.

\item 

What is the effect of the boundary conditions of the strip graph $G_s$ on
${\cal B}$?  For homogeneous strip graphs with (FBC$_y$,FBC$_x$) boundary
conditions, in the cases studied in \cite{strip} where ${\cal B}$ is
nontrivial, it consists of arcs and hence does contain endpoint singularities.
In these cases, the $\lambda_j$'s are algebraic expressions and the arcs
comprising ${\cal B}$ run between branch point zeros of algebraic roots of
polynomials in $q$.  An arc can cross the real axis $q$ axis if and only if it
is self-conjugate.  ${\cal B}$ contains a line segment on the positive real
axis, say in the interval $q_1 < q < q_2$, if there are two $\lambda_j$'s,
e.g., of the form $\lambda_{j,j'}=R(q) \pm \sqrt{S(q)}$ with $R$ and $S$ being
polynomials in $q$ such that $S < 0$ in the interval $q_1 < q < q_2$.

\item

Again concerning the question of the role of boundary conditions:
for the strip graph $(G_s)_m$ with a given type of transverse boundary
conditions BC$_y$, the chromatic polynomial for PBC$_x$ has a larger 
$N_a$ than the chromatic polynomial for FBC$_x$, and the corresponding loci 
${\cal B}$ are different. 

\item

In the examples we have studied, we find that for a given type of strip graph
$G_s$ with FBC$_y$, the chromatic polynomials for PBC$_x$ and TPBC$_x$ boundary
conditions (i.e., cyclic and M\"obius strips) have the same $a_j$, although in
general different $c_j$.  It follows that the loci ${\cal B}$ are the same for
these two different longitudinal boundary condition choices \cite{pg,wcy,pm}.

\item

However, in the case of PBC$_y$, the reversal of orientation involved in going
from PBC$_x$ to TPBC$_x$ longitudinal boundary conditions (i.e. from torus to
Klein bottle topology) can lead to the removal of some of the $a_j$'s that were
present; i.e., $P$ for the (PBC$_y$,TPBC$_x$) strip may involve only a subset
of the $a_j$'s that are present for the (PBC$_y$,PBC$_x$) strip.  For example,
for the $L_y=3$ strips of the square lattice with (PBC$_y$, PBC$_x$) boundary
conditions, there are $N_a=8$ $a_j$'s, but for the strip with (PBC$_y$,
TPBC$_x$) boundary conditions only a subset of $N_a=5$ of these terms occurs in
$P$ \cite{tk}.  None of the three $a_j$'s that are absent from $P$ in the
TPBC$_x$ case is leading, so that ${\cal B}$ is the same for both of these
families.  

\item 

How do $W$ functions in region $R_1$ compare for different boundary conditions?
We have shown that, for a given type of strip graph $G_s$, in the region $R_1$
defined for the PBC$_x$ boundary conditions, the $W$ function is the same for 
FBC$_x$, PBC$_x$, and TPBC$_x$. 

\item

For strips of regular lattices, as one increases $L_y$, how does $W$ approach
the limit for the 2D lattice?  We have found that with $q > q_c$ for a given
lattice type $\Lambda$, the approach of $W$ to its 2D thermodynamic limit as
$L_y$ increases is quite rapid; for moderate values of $q$ and $L_y \simeq 4$,
$W(\Lambda,(L_x=\infty) \times L_y,q)$ is within about 5 \% and ${\cal
O}(10^{-3})$ of the 2D value $W(\Lambda,(L_x=\infty) \times (L_y=\infty),q)$
for FBC$_y$ and PBC$_y$, respectively for a strip graph of the lattice
$\Lambda$.  We have proved that the approach of $W$ to the 2D thermodynamic
limit is monotonic for FBC$_y$ and non-monotonic, although more rapid, for
PBC$_y$ \cite{w2d}.  (By the result noted in the previous item, the $W$
function for these values of $q$, in region $R_1$, is independent of the
longitudinal boundary conditions.)

\item 

Concerning the analytic properties of $W$, one property is that $W$
may have isolated branch point singularities (zeros) for strips with 
(FBC$_y$,FBC$_x$) and (PBC$_y$,FBC$_x$). 
For example, for the (FBC$_y$,FBC$_x$) strips of the square lattice with 
$L_y=2$, and $L_y=3$, the respective $W$ functions have square and cube root
branch point singularities, while for the (PBC$_y$,FBC$_x$) strips of the
square lattice with $L_y=3$, i.e. transverse cross sections forming
triangles, the $W$ functions have cube root branch point singularities 
where they vanish.  Note that for these cases, $R_1$ is the full $q$ plane, and
the continuous locus ${\cal B}=\emptyset$. 
In contrast, for the corresponding lattice strips PBC$_x$, 
although the $W$ functions have the same expressions in 
the respective $R_1$ regions, given below as eqs. (\ref{wsqly2r1}), 
(\ref{wsqly3r1}), and (\ref{wsqly3pbc}), 
the zeros of the expressions do not occur in the 
$R_1$ regions where these expressions apply, and so the $W$ functions in these
$R_1$ regions do not have any isolated branch point singularities.  We have 
also found this to be true of other strips. 

\end{enumerate}	

\section{Chromatic Polynomials and Loci ${\cal B}$ for Some Specific Families}

Here we present exact calculations of chromatic polynomials, $W$ functions, and
loci ${\cal B}$ for various families of graphs.

\subsection{Strips with Free Transverse and Longitudinal Boundary Conditions}

We first show ${\cal B}$ for the strip graphs of the square lattice with 
(FBC$_y$,FBC$_x$) and $L_y=2,3$.  The relevant generating functions are given
in \cite{strip}.  

\begin{figure}
\begin{center}
\leavevmode
\epsfxsize=3.5in
\epsffile{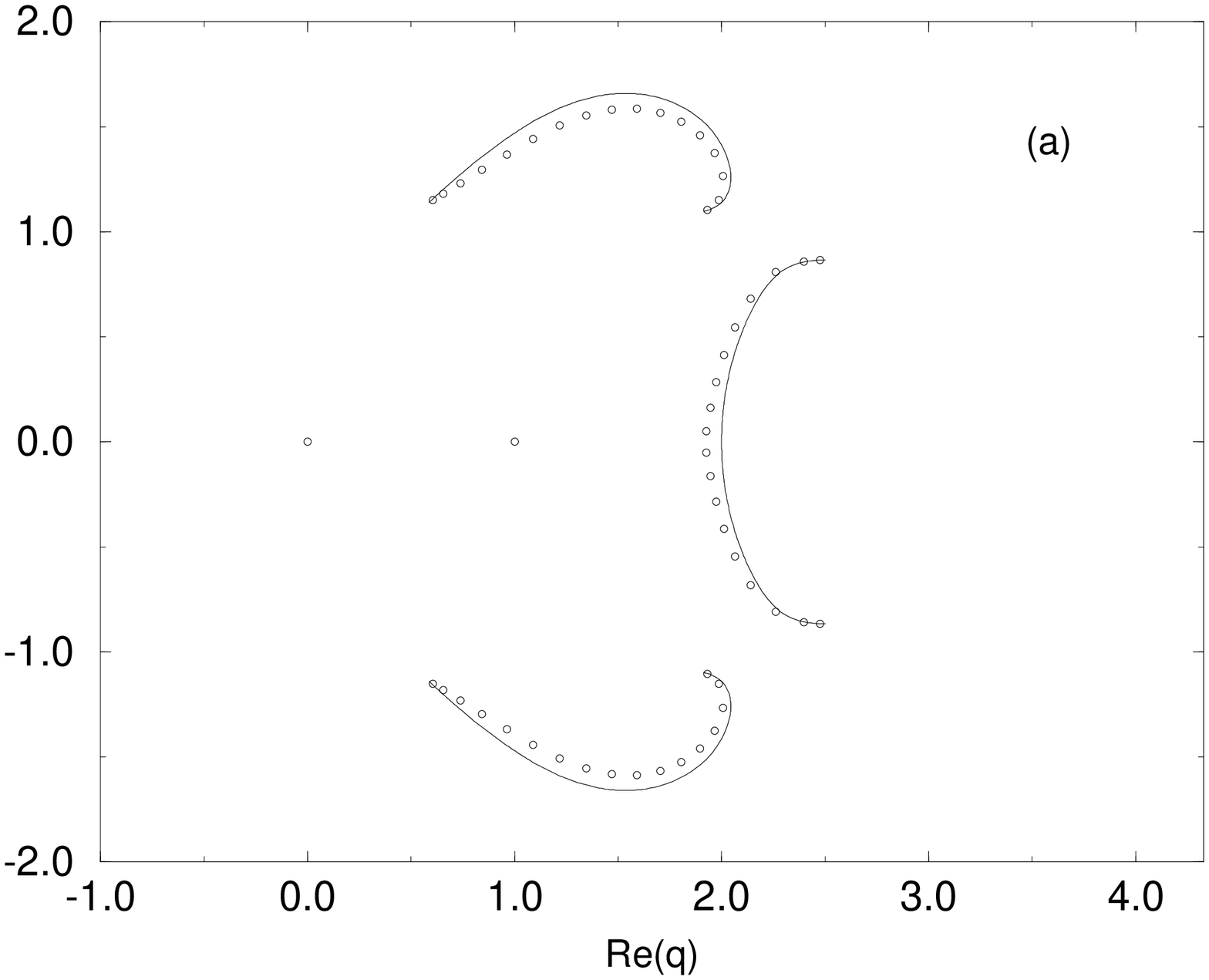}
\end{center}
\vspace{-4cm}
\begin{center}
\leavevmode
\epsfxsize=3.5in
\epsffile{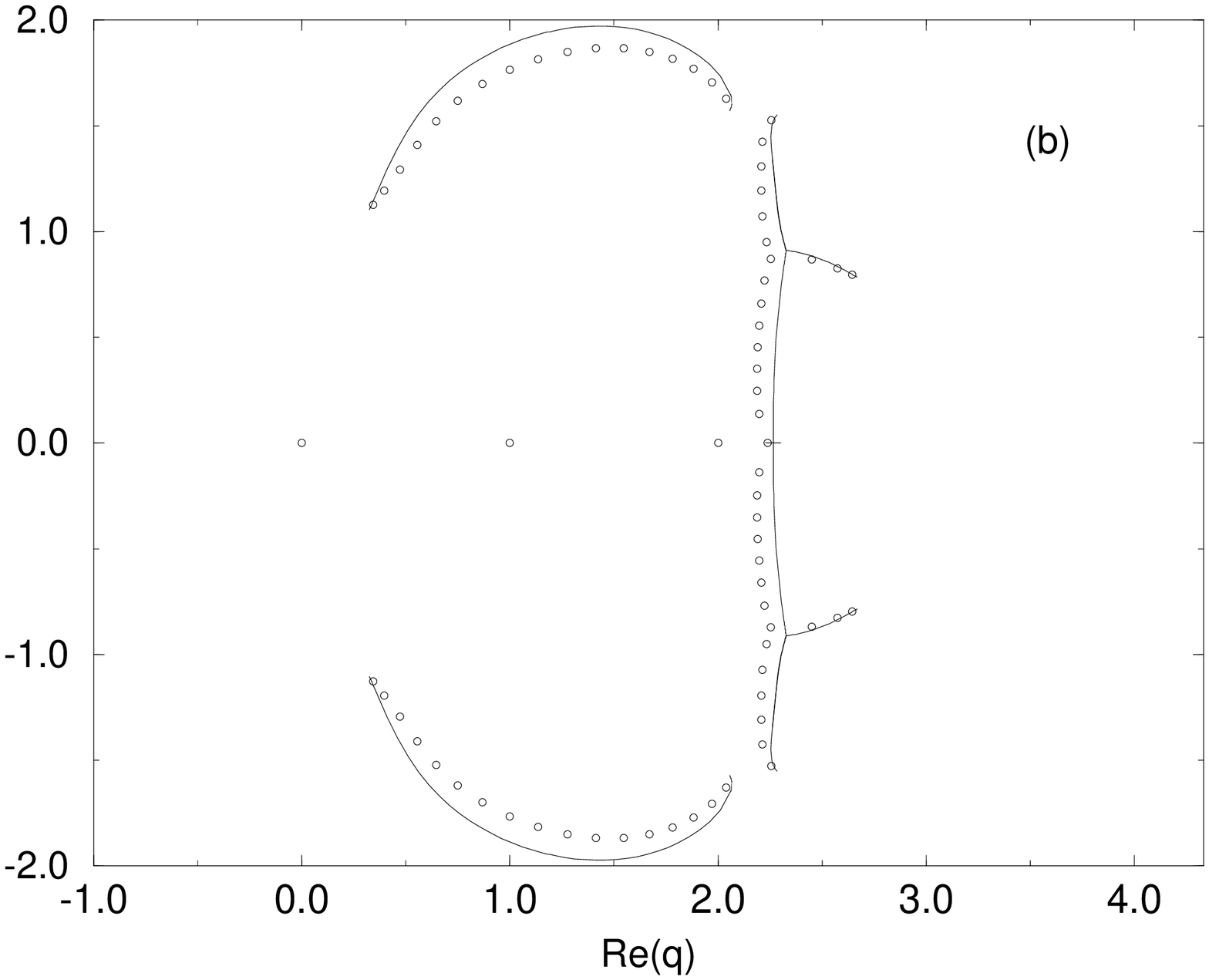}
\end{center}
\caption{\footnotesize{ ${\cal B}$ for $W(\{G_{sq(L_y)}\},q)$ with 
$L_y =$ (a) 3, (b) 4, where $\{G_{sq(L_y)}\}$ denotes the
$(L_x=\infty) \times L_y$ strip of the square lattice.
For comparison, the zeros of the chromatic polynomial
$P((G_{sq(L_y)})_m,q)$ for (a) $L_y=3$, $m=16$ (hence $n=54$ vertices) and
(b) $L_y=4$, $m=16$ (hence $n=72$) are shown.}}
\label{sqstrip}
\end{figure}

\subsection{Strips with Free Transverse and Periodic or M\"obius Longitudinal
Boundary Conditions} 

\begin{figure}
\begin{center}
\leavevmode
\epsfxsize=3.5in
\epsffile{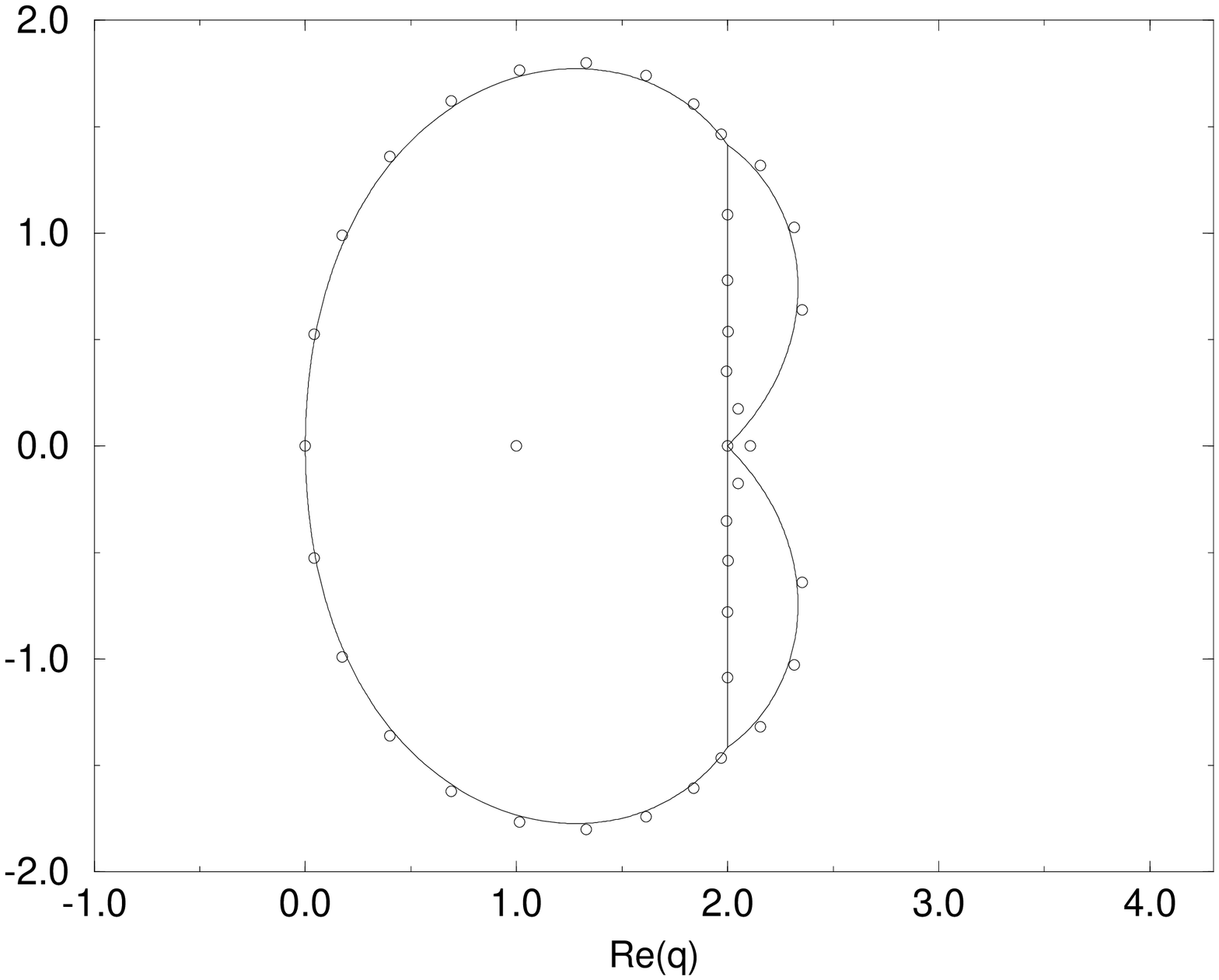}
\end{center}
\vspace{-4cm}
\begin{center}
\leavevmode
\epsfxsize=3.5in
\epsffile{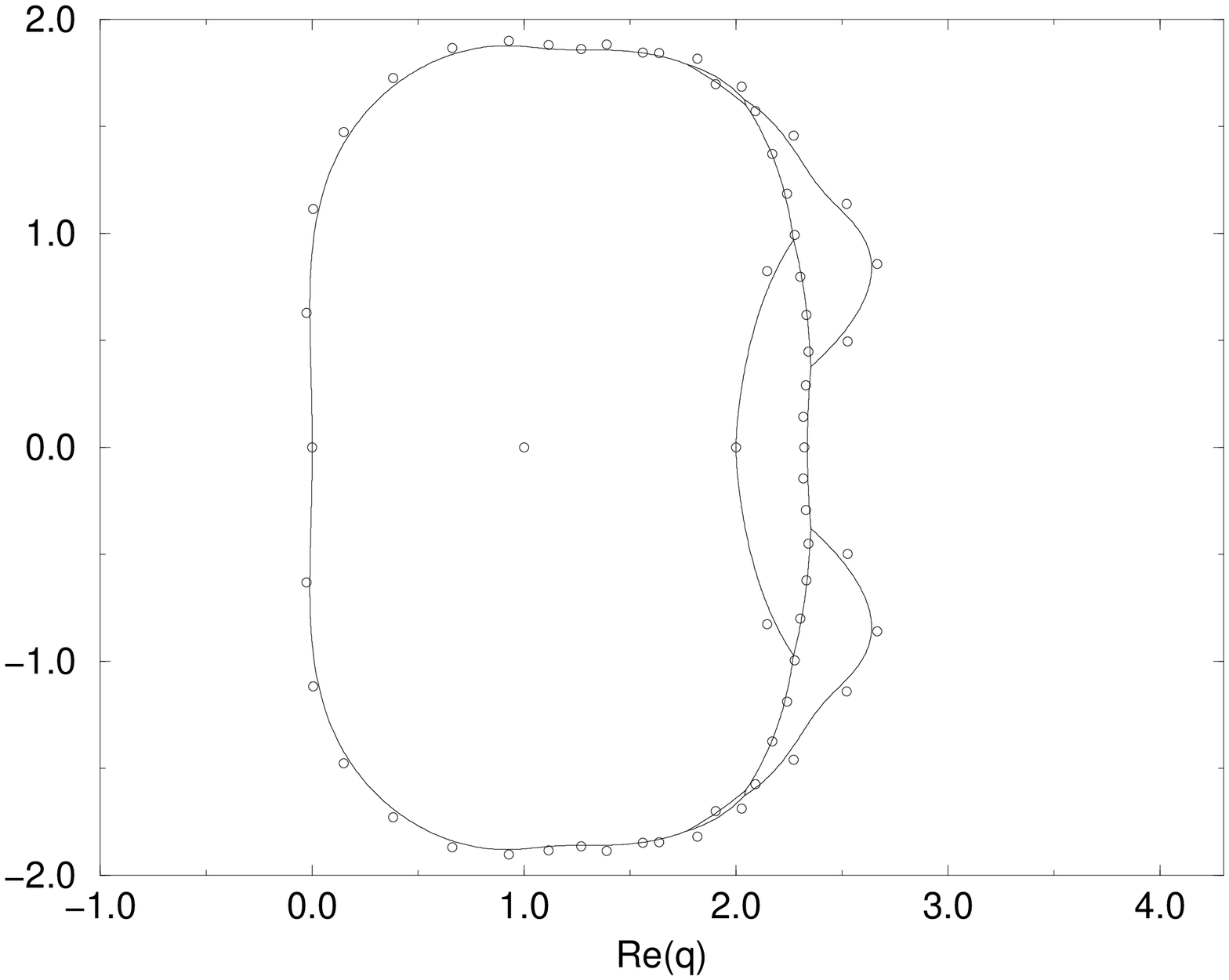}
\end{center}
\caption{\footnotesize{ ${\cal B}$ for the infinite-length limit of strips of
the square lattice with (FBC$_y$,PBC$_x$) (cyclic) boundary conditions and 
$L_y =$ (a) 2, (b) 3. For comparison, the zeros of the chromatic polynomial
$P((G_{sq(L_y)})_m,q)$ for (a) $m=19$ (hence $n=38$) and
(b) $m=20$ (hence $n=60$) are shown for comparison.  The ${\cal B}$'s for the
corresponding strips with (FBC$_y$,TPBC$_x$) (M\"obius) boundary conditions are
identical.}}
\label{cycsqstrip}
\end{figure}

For definiteness, we consider cyclic and M\"obius strips of the square 
lattice.  For $L_y=1$, these reduce to the circuit graph, for which 
${\cal B}$ is the circle $|q-1|=1$.  For the $L_y=2$ strips of this type, from
the $P$ in \cite{bds} we obtain ${\cal B}$ given in Fig. \ref{cycsqstrip}. 

For the cyclic $L_y=3$ strip of the square lattice, we find \cite{wcy}
\beqs
& & P(sq(L_y = 3,cyc.)_m,q) = (q^3-5q^2+6q-1)(-1)^m \cr\cr
& & + (q^2-3q+1)\Bigl [(q-1)^m+(q-2)^m+(q-4)^m \Bigr ]
+ (q-1)[-(q-2)^2]^m \cr\cr
& & + \Bigl [ (\lambda_{sq,6})^m+(\lambda_{sq,7})^m \Bigr ]
+(q-1)\Bigl [(\lambda_{sq,8})^m+(\lambda_{sq,9})^m+(\lambda_{sq,10})^m \Bigr ]
\label{psqly3}
\eeqs
where
\beq
\lambda_{sq,(6,7)} = \frac{1}{2}\Biggl [ (q-2)(q^2-3q+5) \pm
\Bigl \{ (q^2-5q+7)(q^4-5q^3+11q^2-12q+8) \Bigr \}^{1/2} \Biggr ]
\label{lambda67}
\eeq
and $\lambda_{sq,j}$, $j=8,9,10$, are the roots of the cubic equation
\beq
\xi^3+b_{sq,21}\xi^2+b_{sq,22}\xi+b_{sq,23}=0
\label{sqcubic}
\eeq
with
\beq
b_{sq,21}=2q^2-9q+12
\label{bsq21}
\eeq
\beq
b_{sq,22}=q^4-10q^3+36q^2-56q+31
\eeq
\label{bsq22}
\beq
b_{sq,23}=-(q-1)(q^4-9q^3+29q^2-40q+22) \ .
\label{bsq23}
\eeq
and for the corresponding M\"obius strip \cite{pm} 
\beqs
& & P(sq(L_y = 3,\ Mb.)_m,q) = (q^2-3q+1)(-1)^m
-(q-1)^m+(q-2)^m-(q-4)^m \cr\cr & & -(q-1)[-(q-2)^2]^m + \Bigl [
(\lambda_{sq,6})^m+(\lambda_{sq,7})^m \Bigr ] +(q-1)\Bigl
[(\lambda_{sq,8})^m+(\lambda_{sq,9})^m+(\lambda_{sq,10})^m \Bigr ] \cr\cr &&
\label{psqly3tw}
\eeqs

For the ($L_x \to \infty$ limits of the) $L_y=1$ and $L_y=2$ cyclic strips,
$q_c=2$, while for the $L_y=3$ cyclic strip, $q_c=2.33654$.  These values are
obtained from exact solutions for the respective $W$ and ${\cal B}$ for these
strips \cite{w,wcy,pm}.  For these cyclic graphs this point
$q_c$ is a non-decreasing function of $L_y$.  In the limit $L_y \to \infty$,
continuity arguments imply that $q_c$ approaches the value for the 2D square
lattice, which is $q_c(sq)=3$ \cite{lieb}.

The $W$ functions for the regions are given in \cite{w,wcy,pm}.  In particular,
note that in the respective region $R_1$ for each family, 
\beq
W(sq(L_y=1),cyc.,q)=q-1
\label{wsqly1r1}
\eeq
and, for both cyclic and M\"obius strips, 
\beq
W(sq(L_y=2),cyc.,q)=(q^2-3q+3)^{1/2}
\label{wsqly2r1}
\eeq
\beqs
W(sq(L_y=3,cyc.),q) & = & 2^{-1/3}\biggl [ (q-2)(q^2-3q+5) + \nonumber \\ & &
\Bigl [(q^2-5q+7)(q^4-5q^3+11q^2-12q+8) \Bigr ]^{1/2} \biggr ]^{1/3}
\label{wsqly3r1}
\eeqs
These $W$ functions are the same as for the corresponding strips with 
(FBC$_y$,PBC$_x$), which is a general result, as noted above. 

\subsection{Homeomorphic Expansion of Cyclic and M\"obius Strips}

\begin{figure}
\vspace{-4cm}
\centering
\leavevmode
\epsfxsize=3.0in
\begin{center}
\leavevmode
\epsffile{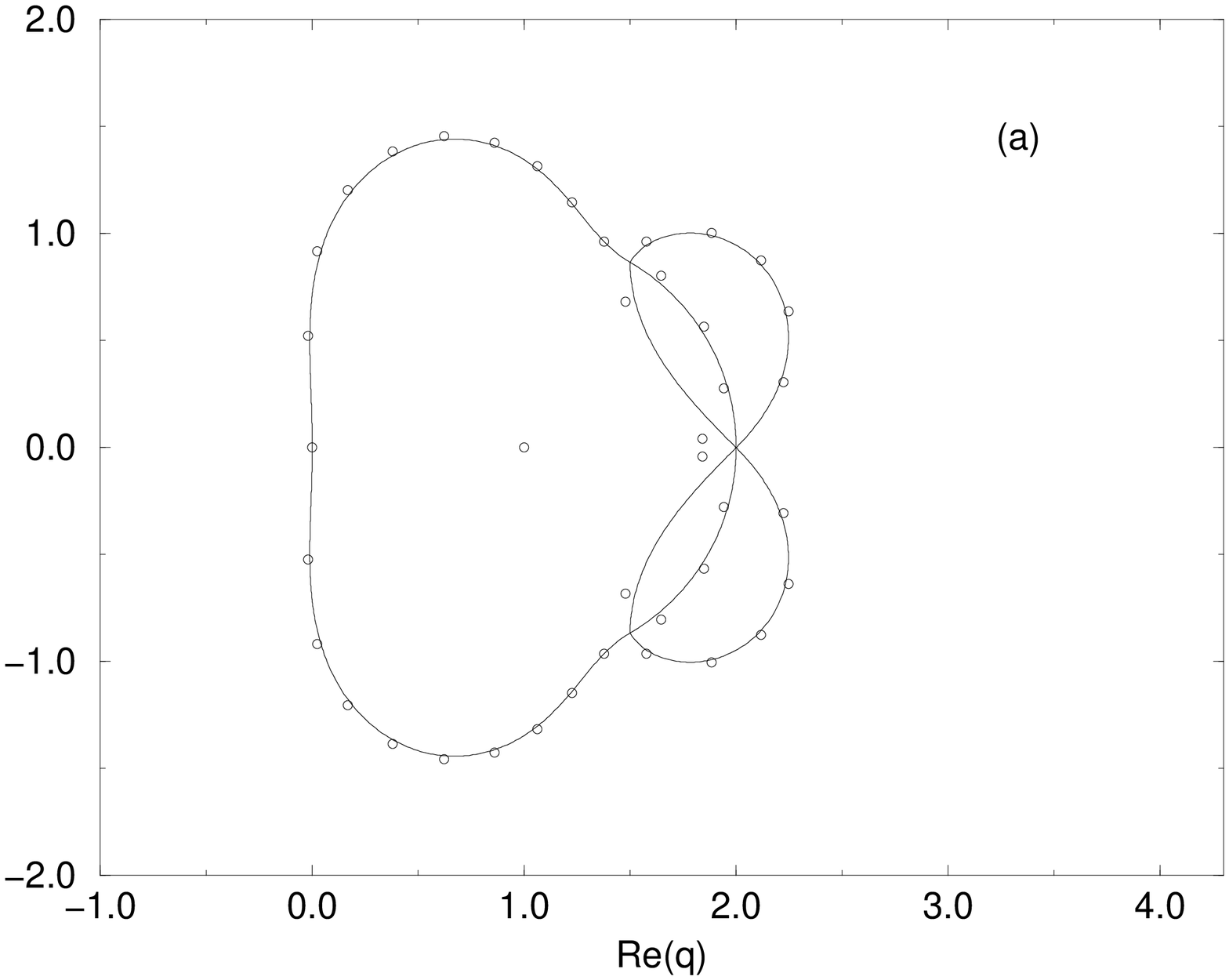}
\end{center}
\vspace{-2cm}
\begin{center}
\leavevmode
\epsfxsize=3.0in
\epsffile{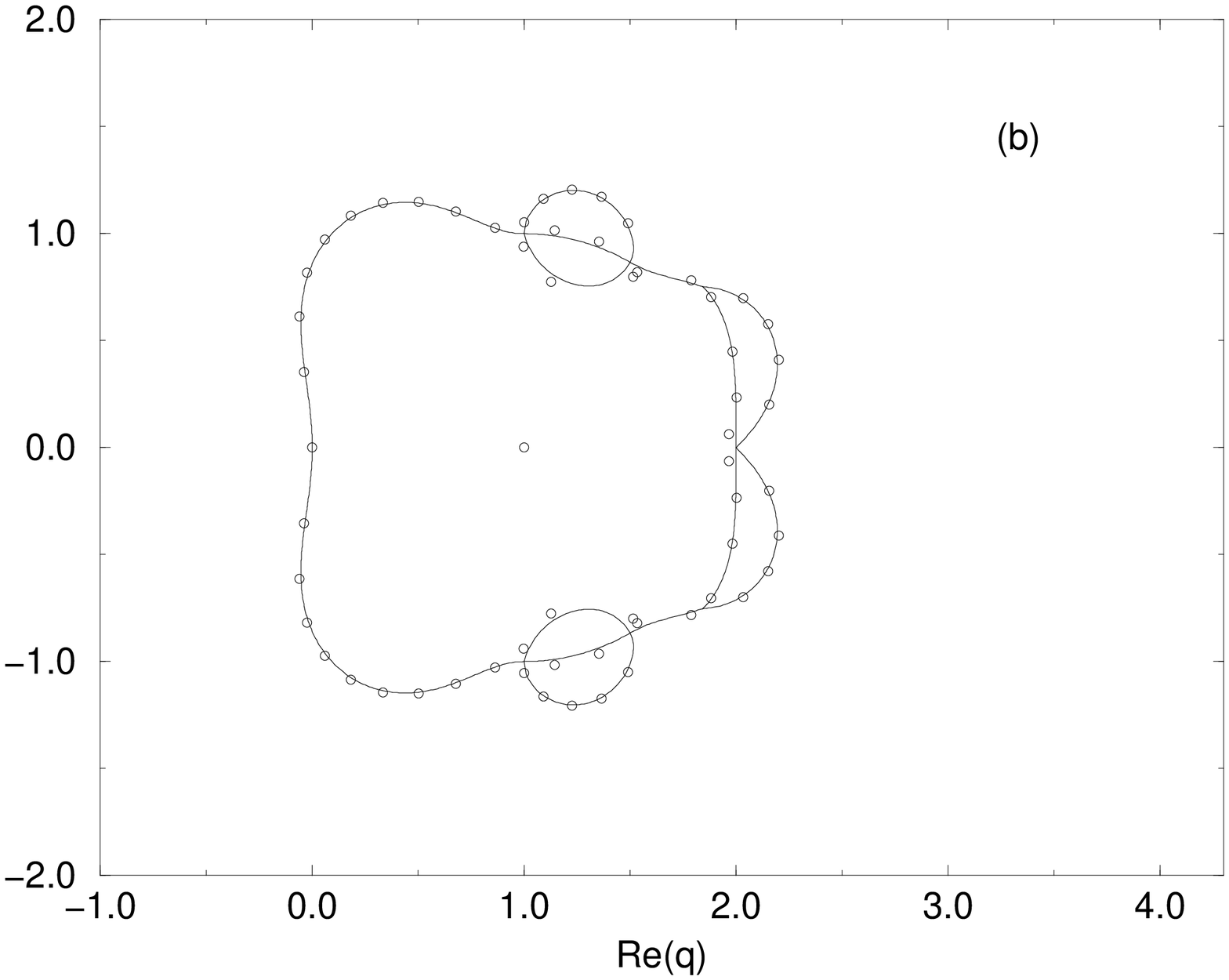}
\end{center}
\vspace{-2cm}
\caption{\footnotesize{${\cal B}$ for $\lim_{m \to \infty}(Ch)_{k,m,cyc.}$
with $k=$ (a) 3 (b) 4. Chromatic zeros are shown for the cyclic case with
$m=10$, i.e., $n=$ (a) 40 (b) 60.}}
\label{homsq}
\end{figure}

For comparison with the previous figure, we show homeomorphic expansions
obtained by starting with cyclic and M\"obius $L_y=2$ square strips consisting
of $m$ squares, and inserting $k-2$ additional vertices on each
horizontal edge.  We denote these graphs as as $(Ch)_{k,m,cyc.}$ and
$(Ch)_{k,m,Mb.}$.  These graphs can be regarded as cyclic and M\"obius strips
of $m$ $p$-sided polygons, where $p=2k$, such that each $p$-gon intersects the
previous one on one of its edges, and intersects the next one on its opposite
edge.  It is convenient to define
\beq
D_k(q) = \frac{P(C_k,q)}{q(q-1)} = 
\sum_{s=0}^{k-2}(-1)^s {{k-1}\choose {s}} q^{k-2-s}
\label{dk}
\eeq
where
$P(C_m,q)$ is the chromatic polynomial for the circuit
(cyclic) graph $C_m$ with $m$ vertices, $P(C_m,q) = (q-1)^m + (q-1)(-1)^m$. 
We calculate \cite{pg} 
\beqs
P((Ch)_{k,m,cyc.},q) & = & q^2-3q+1+ (D_{2k})^m + \cr\cr
& & (q-1)\biggl [ 
\Bigl ( (-1)^{k+1}D_{k+1}+D_k \Bigr )^m + 
\Bigl ( (-1)^{k+1}D_{k+1}-D_k \Bigr )^m \biggr ]
\label{pg1}
\eeqs
\beqs
P((Ch)_{k,m,cyc.,Mb.},q) & = & -1 + (D_{2k})^m + \cr\cr
& & (-1)^k(q-1)\biggl [
\Bigl ( (-1)^{k+1}D_{k+1}+D_k \Bigr )^m - 
\Bigl ( (-1)^{k+1}D_{k+1}-D_k \Bigr )^m \biggr ] 
\label{pchtw}
\eeqs
In Fig. \ref{homsq} we show ${\cal B}$ and chromatic zeros for the cases 
$k=3,4$.  In region $R_1$, which includes the real axis for $q \ge 2$, 
\beq
W = (D_{2k})^{\frac{1}{2(k-1)}}
\label{wr1pg}
\eeq

\newpage

\subsection{Cyclic Chains of Polygons}

Consider a cyclic chain composed of $m$
subunits, each subunit consisting of a $p$-sided polygon with one of its
vertices attached to a line segment of length $e_g$ edges (bonds).  Thus, the
members of each successive pair of polygons are separated from each other by a
distance (gap) of $e_g$ bonds along these line segments, with $e_g=0$
representing the case of contiguous polygons.  Since each polygon is connected
to the rest of the chain at two vertices (taken to be at the same relative
positions on the polygons in all cases), this family of graphs depends on two
additional parameters, namely the number of edges of the polygons between these
two connection vertices, moving in opposite directions along the polygon, $e_1$
and $e_2$. We denote this family of cyclic polygon chain graphs as
$G_{e_1,e_2,e_g,m;o}$ and define, in this context, $p=e_1+e_2$.  
A symmetry property is 
\beq
P(G_{e_1,e_2,e_g,m},q) = P(G_{e_2,e_1,e_g,m},q)
\label{ppcsym}
\eeq
We calculate \cite{nec} 
\beq
P(G_{e_1,e_2,e_g,m},q) = (a_1)^m + (q-1)(a_2)^m
\label{ppc}
\eeq
where
\beq
a_1 = (q-1)^{e_g+1}D_p
\label{a1}
\eeq
\beqs
a_2 & = & (-1)^{p+e_g}q^{-1}\Bigl [ q-2  + (1-q)^{e_1} + (1-q)^{e_2} \Bigr ] \\
& = &  (-1)^{p+e_g}\biggl [ 1-p - \sum_{s=2}^{e_1}{e_1 \choose s}(-q)^{s-1}
- \sum_{s=2}^{e_2}{e_2 \choose s}(-q)^{s-1} \biggr ]
\label{a2}
\eeqs

\begin{figure}
\centering
\leavevmode
\epsfxsize=3.5in
\begin{center}
\leavevmode
\epsffile{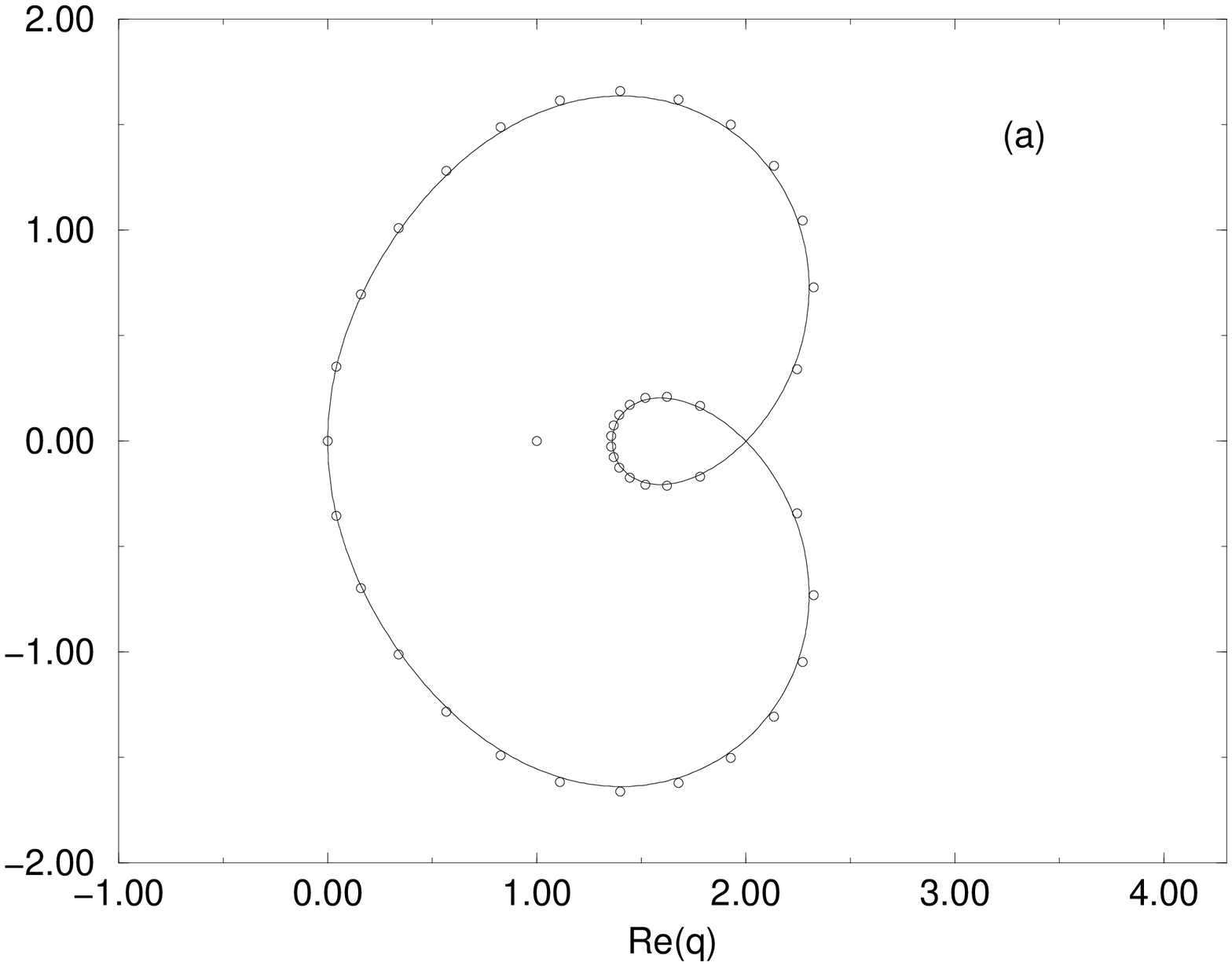}
\end{center}
\vspace{-4cm}
\begin{center}
\leavevmode
\epsfxsize=3.5in
\epsffile{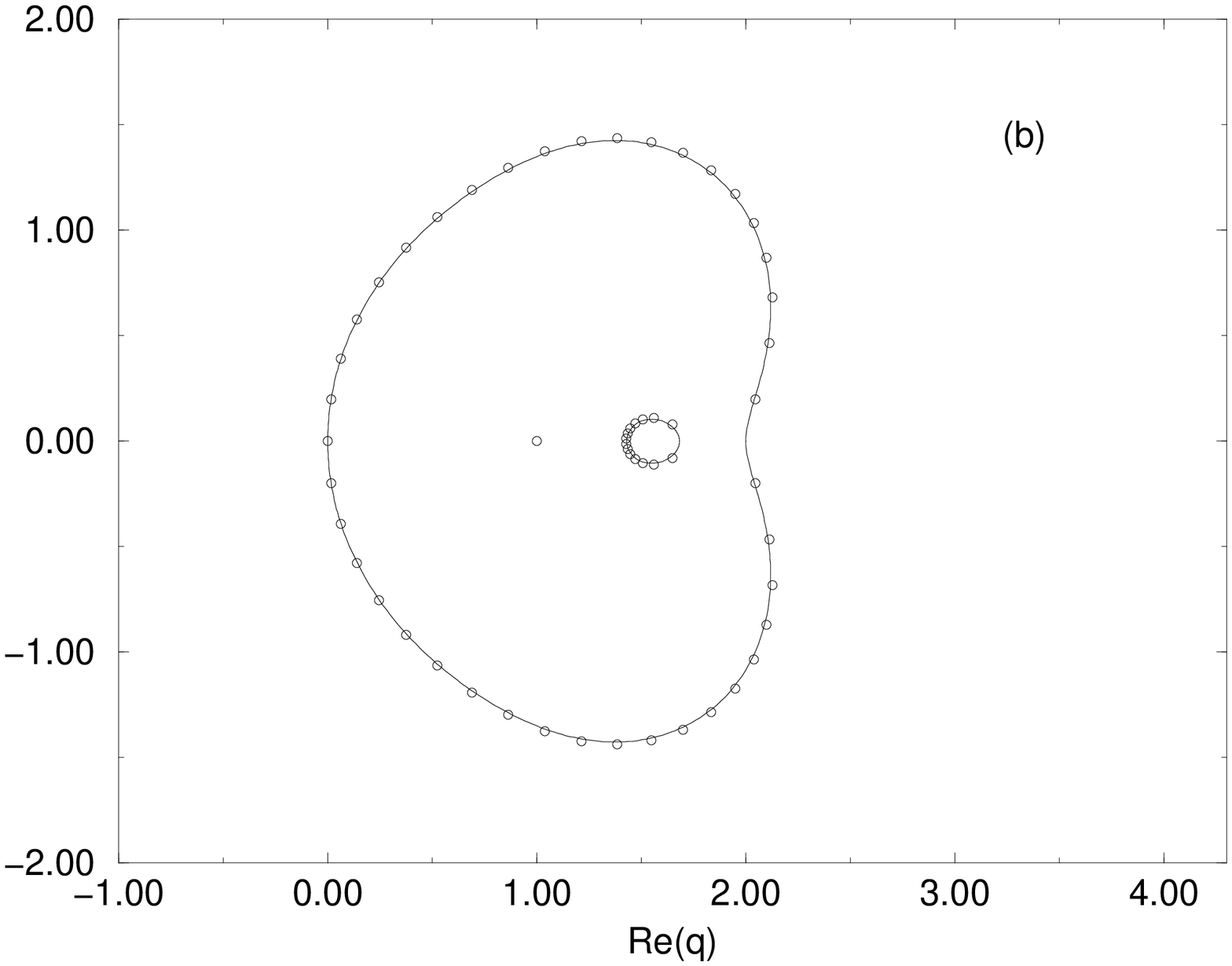}
\end{center}
\vspace{-2cm}
\caption{\footnotesize{Boundary ${\cal B}$ in the $q$ plane for $W$ function
for $\lim_{m \to \infty} G_{e_1,e_2,e_g,m}$ with $(e_1,e_2,e_g)=$
(a) (2,2,0), (b) (2,2,1). Chromatic zeros for $m=14$ (i.e., $n=42$ and
$n=56$ for (a) and (b)) are shown for comparison.}}
\label{nec22g01}
\end{figure}

The locus ${\cal B}$ obtained by taking the limit $m \to \infty$ is shown in
Fig. \ref{nec22g01} for the cases $(e_1,e_2,e_g)=$ (a) (2,2,0), (b)
(2,2,1). One sees the interesting phenomenon that as the number of edges in the
gap $e_g$ increases from 0, the multiple point that was present on ${\cal B}$
for $e_g=0$ disappears and ${\cal B}$ decomposes into two separate components.
As $e_g$ increases further, the inner boundary shrinks monotonically around the
point at its center.  This family also illustrates the fact that one can take
the limit $n \to \infty$ in a different way, letting $e_1$, $e_2$, or $e_g$ go
to infinity with $m$ held fixed.  For the nontrivial case where $min(e_1,e_2) >
1$, we find that, if $p$ is even, then $q_c=2$ while if $p$ is odd, then 
$q_c < 2$ and, for fixed $(e_1,e_2)$, $q_c$ increases monotonically as $e_g$
increases, approaching 2 from below as $e_g \to \infty$. In region $R_1$,
\beq
W = [(q-1)^{e_g+1}D_p]^{\frac{1}{p+e_g-1}}
\label{wr1}
\eeq
which is again the same $W$ function as for the corresponding polygon chain
graph with free (open) longitudinal boundary conditions.

\subsection{Families with Periodic Transverse and Free Longitudinal 
Boundary Conditions}

Here we consider strip graphs with (PBC$_y$,FBC$_x$). 
For the strip graph of the square lattice with $L_y=3$, i.e., cross sections
forming triangles, and FBC$_x$, we find \cite{strip,w2d}
\beq
P(sq(L_y=3)_m,PBC_y,FBC_x,q) = q(q-1)(q-2)(q^3-6q^2+14q-13)^{m-1}
\label{psqly3pbcy}
\eeq
whence
\beq
W(sq(L_y=3),PBC_y,FBC_x,q) = (q^3-6q^2+14q-13)^{1/3}
\label{wsqly3pbc}
\eeq 
with ${\cal B}=\emptyset$. Results for this lattice strip with larger
$L_y=4$, i.e. cross sections forming squares, and for the triangular lattice, 
are given in \cite{w2d}. 

\subsection{Families with Periodic Transverse and (Twisted) Periodic 
Longitudinal Boundary Conditions}

With N. Biggs, we have recently studied the case of the strip of the square
lattice with (PBC$_y$,PBC$_x$) (torus) and (PBC$_y$,TPBC$_x$) (Klein bottle)
boundary conditions; see Ref. \cite{tk} for the chromatic polynomials and $W$
functions.  We find $q_c=3$, which, interestingly, is the same as for the full
2D square lattice.

\subsection{Families with Noncompact ${\cal B}$}

As an example of a family of graphs with a noncompact ${\cal B}$ is obtained as
follows \cite{wa}.  We start with the join $(K_p) + G_r$ and remove some bonds
from $K_p$.  The simplest case is to let $G=\overline K_r$, $K_p=K_2$, and
remove the edge connecting the two vertices of the $K_2$.  We can
homeomorphically expand this by adding degree-2 vertices to the $r$ edges
connecting each of the two vertices of the original $K_2$ to the $r$ vertices
of the $\overline K_r$.  We denote this family as $H_{k,r}$ \cite{wa2,wa3}.  We
find 
\cite{wa2,wa3}
\beqs 
P(H_{k,r},q) & = & q(q-1)\biggl [ D_{2k-2}(D_k)^{r-2} 
-(q-1)(D_{k-1})^2 \Bigl [(D_k)^{r-2} - \bigl \{ (q-1)D_{k-1} \bigr \}^{r-2}
\Bigr ] \biggr ] \cr\cr 
& & = q(q-1)\biggl [ (q-1)^{r-1}(D_{k-1})^r + (D_k)^r
\biggr ]
\label{phkr}
\eeqs

\begin{figure}
\vspace{-4cm}
\centering
\leavevmode
\epsfxsize=3.5in
\begin{center}
\leavevmode
\epsffile{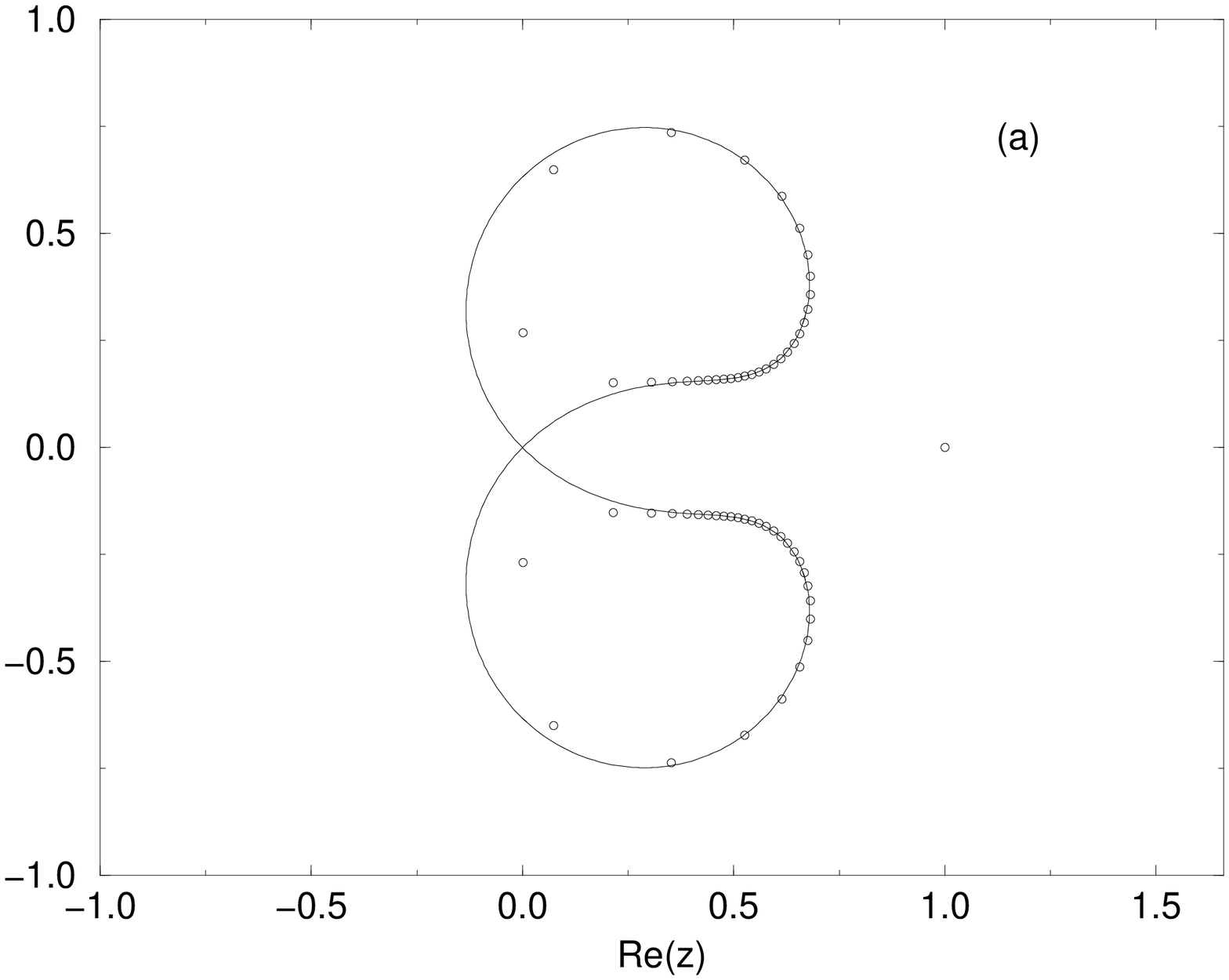}
\end{center}
\vspace{-4cm}
\begin{center}
\leavevmode
\epsfxsize=3.5in
\epsffile{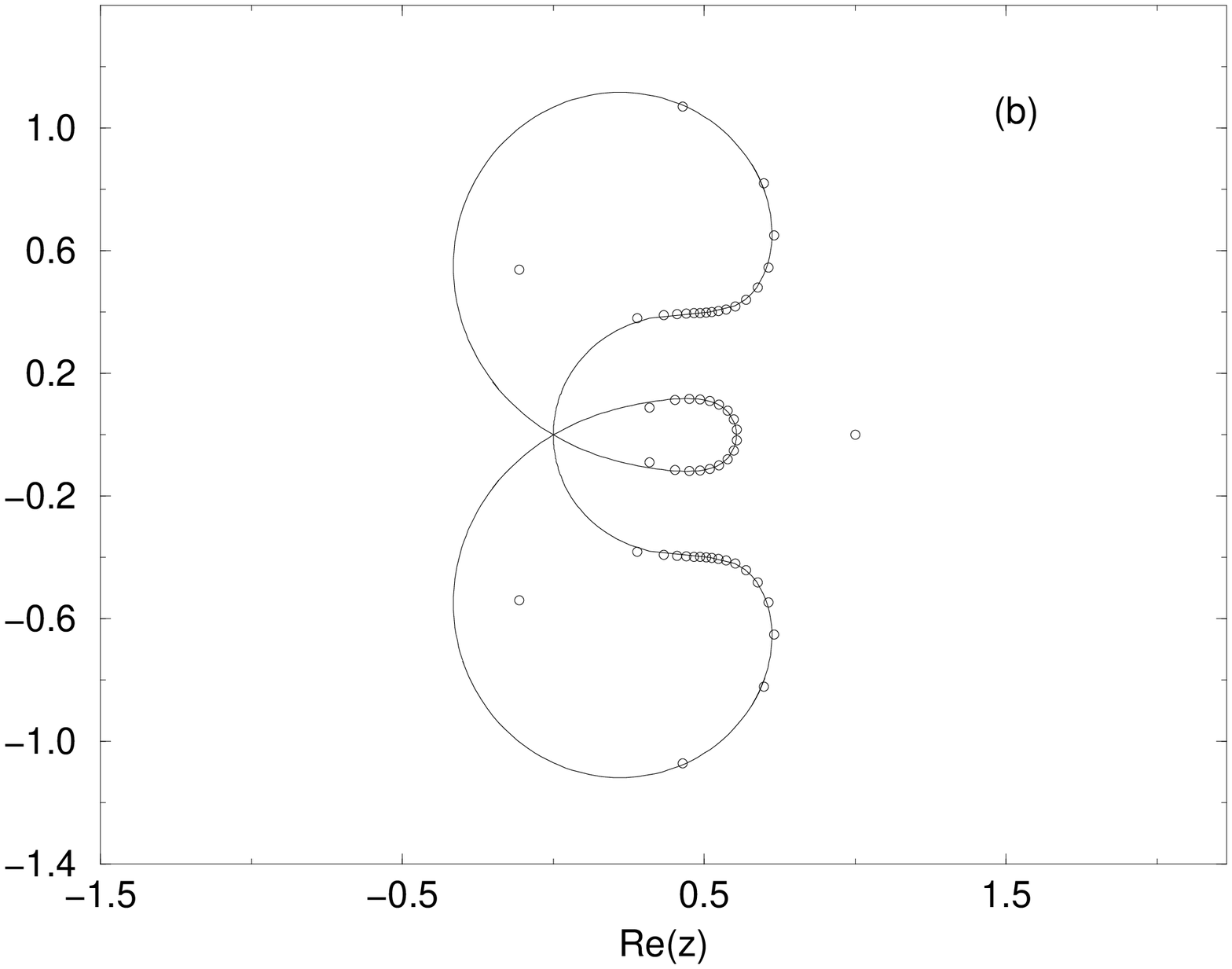}
\end{center}
\vspace{-2cm}
\caption{\footnotesize{Boundary ${\cal B}$ in the $z=1/q$ plane for
$\lim_{r \to \infty}H_{k,r}$ with $k=$ (a) 4 (b) 5.
Chromatic zeros for $H_{k,r}$ with $(k,r)=$ (a) (4,30) (b) (5,18)
are shown for comparison.}}
\label{hamboundaryk4z}
\end{figure}

Because of the noncompactness of ${\cal B}$ in the $q$ plane, it is more
convenient to plot it in the $z=1/q$ plane.  In Fig. \ref{hamboundaryk4z} we
show two typical plots.  For further details, see Refs. \cite{wa,wa3,wa2}. 
General bounds on chromatic zeros of graph have been discussed in a number of
papers, e.g., \cite{bds} and recently \cite{sokal}.

\section{Zeros and their Accumulation Sets for Potts Partition Functions at 
Finite Temperature or Tutte Dichromatic Polynomials}

We have recently generalized our studies to the case of the $q$-state Potts
model at arbitrary temperatures.  At temperature $T$ on a graph $G$ or 
lattice $\Lambda$ this model is defined by the partition function 
\beq 
Z = \sum_{ \{ \sigma_n \} } e^{-\beta {\cal H}}
\label{zfun}
\eeq
with the Hamiltonian
\beq
{\cal H} = -J \sum_{\langle i j \rangle} \delta_{\sigma_i \sigma_j}
\label{ham}
\eeq
where $\sigma_i=1,...,q$ are the spin variables on each vertex $i \in G$; 
$\beta = (k_BT)^{-1}$; and $\langle i j \rangle$
denotes pairs of adjacent vertices.  (Our terminology for this partition 
function is standard in physics; obviously $Z$ should not be confused with 
the partition function $p(n)$ in number theory and combinatorics.) 
We use the notation $K = \beta J$, 
\beq
a = u^{-1} = e^{K}
\label{a}
\eeq
and 
\beq
v = a-1
\label{v}
\eeq
and denote the (reduced) free energy per vertex (site) as
$f = -\beta F = \lim_{n \to \infty} n^{-1} \ln Z$.
The partition function can be written as 
\beq
Z(G,q,a) = \sum_{\{\sigma_i \}} \prod_{\langle i j \rangle} 
(1+v\delta_{\sigma_i,\sigma_j})
\label{zv}
\eeq
which shows that it is a polynomial in $q$ and $v$ or $a$. From (\ref{zv}), it
follows that \cite{whit,tutte1,kf,wurev}
\beq
Z(G,q,a)=\sum_{G^\prime \subseteq G} q^{k(G^\prime)}v^{e(G^\prime)}
\label{tkf}
\eeq
where $G^\prime$ is a subgraph of $G$ and $e(G^\prime)$ and $k(G^\prime)$ 
denote the edges (bonds) and connected components, including single vertices, 
of $G^\prime$, respectively. 

The ferromagnetic and 
antiferromagnetic signs of the spin-spin exchange coupling are $J > 0$ and 
$J < 0$, respectively, and hence as the temperature varies from 0 to $\infty$,
the variable $a$ varies from 0 to 1 for the Potts antiferromagnet and from 
$\infty$ to 1 for the Potts ferromagnet.  At $T=\infty$, i.e., $\beta=0$ or 
$a=1$, the Potts ferromagnet and antiferromagnet are identical, since $J$ 
does not enter in $Z$, which reduces simply to 
\beq
Z(G,q,a=1)=q^{n(G)}
\label{za1}
\eeq
As indicated in eq. (\ref{pz}), 
the $T=0$ ($a=0$) case yields the chromatic polynomial. For other special cases
one has the elementary results $Z(G,q=0,a)=0$ and 
\beq
Z(G,q=1,a)=a^{e(G)}
\label{zq1}
\eeq
The zeros of $Z$ are
now functions of both complex $q$ and $a$.  In the limit $n \to \infty$, 
these zeros have an accumulation set which is a singular submanifold ${\cal B}$
in the ${\mathbb C}^2$ space defined by $q$ and $a$ (or equivalently, $u$). 
Just as, in the one-variable situation, the $W(\{G\},q)$ function was 
singular across the locus ${\cal B}$ in the $q$ plane as it switched its 
analytic form due to a change in the dominant term $a_j(q)$ in (\ref{pgsum}), 
so in the two-variable situation, the reduced free energy is singular across
the locus ${\cal B}$ in the $(q,a)$ space as it switches it analytic form due
to a change in the dominant term $\lambda_j$ in (\ref{zgsum}). 
We denote the slices of this submanifold ${\cal B}$ in the $q$ plane, and in 
the $a$ or $u$ variable as ${\cal B}_q$, ${\cal B}_a$, and ${\cal B}_u$. 

We recall that the partition function $Z(G,q,a)$ is related to the Tutte 
(dichromatic) polynomial $T(G,x,y)$ \cite{tutte1,tutte2} and the rank 
function $R(G,\xi,\eta)$ \cite{whit}.  Defining 
\beq
x=1+\frac{q}{v}
\label{xdef}
\eeq
and
\beq
y=a=v+1
\label{ydef}
\eeq
so that $q=(x-1)(y-1)$, one has 
\beq
T(G,x,y) = (x-1)^{-k(G)}(y-1)^{-n(G)}Z(G,q,v)
\label{tuttez}
\eeq
This is equivalent to the usual expression of the Tutte polynomial in terms of
spanning trees \cite{tutte1}.  The connection with the rank function follows
from the relation $T(G,x+1,y+1)=x^{n-1}R(G,x^{-1},y)$, viz., 
\beq
Z(G,q,v)=q^{n(G)} R(G,\xi=\frac{v}{q},\eta=v)
\label{zwhit}
\eeq

 The partition function or Tutte polynomial can be calculated either by the
iterative use of the deletion-contraction theorem or by a generalization of the
usual transfer matrix method from statistical mechanics. 
For recursive families, from the transfer matrix method, it follows that $Z$ 
has the structure
\beq
Z(G,q,a) = \sum_{j=1}^{N_\lambda} c_j (\lambda_j)^m
\label{zgsum}
\eeq
where the $c_j$ and the $\lambda_j$ depend on $G$ but are independent 
of $m$.  This is a generalization of eq. (\ref{pgsum}) (with the equivalent 
notation $N_\lambda \equiv N_a$ and $\lambda_j \equiv a_j$). 
Given the formula (\ref{tkf}), the zero-temperature limit of the 
Potts antiferromagnet is studied by taking $a \to 0$.  For the Potts
ferromagnet, since $a \to \infty$ as $T \to 0$ and $Z$ diverges like 
$a^{e(G)}$ in this limit, it 
is convenient to use the low-temperature variable $u=1/a=e^{-K}$ and the
reduced partition function $Z_r$ defined by 
\beq
Z_r=a^{-e(G)}Z=u^{e(G)}Z
\label{zr}
\eeq
which has the finite limit $Z_r \to 1$ as $T \to 0$.  For a general strip 
graph $(G_s)_m$ of type $G_s$ and length $L_x=m$, we can write 
\beqs
Z_r((G_s)_m,q,a) & = & u^{e((G_s)_m)}\sum_{j=1}^{N_\lambda} c_j
(\lambda_j)^m \equiv \sum_{j=1}^{N_\lambda} c_j (\lambda_{j,r})^m
\label{zu}
\eeqs
with 
\beq
\lambda_{j,r}=u^{e((G_s)_m)/m}\lambda_j
\label{lamu}
\eeq

An elementary calculation for the circuit graph $C_n$ 
(e.g., \cite{is1d}) yields $Z(G=C_n,q,v) = (q+v)^n + (q-1)v^n$. 
The locus ${\cal B}$ is the solution of the degeneracy equation 
$|a+q-1|=|a-1|$ 
${\cal B}$. The slice in the $q$ plane, ${\cal B}_q$, thus consists of the
circle centered at $q=1-a$ with radius $|1-a|$: 
\beq
q = (1-a)(1 + e^{i\theta}) \ , \quad \theta \in [0,2\pi)
\label{qcirca}
\eeq
For the Potts antiferromagnet, as $a$ increases from 0 to 1, this circle 
contracts monotonically toward the origin, and at $a=1$ it degenerates into a 
point at $q=0$.  For the Potts ferromagnet, as $a$ increases from 1 to
$\infty$, the circle expands into the $Re(q) < 0$ half-plane.  One can also
consider other values of $a$ that do not correspond to physical temperature in
the Potts ferromagnet or antiferromagnet.  In all cases, from
eq. (\ref{qcirca}) it is evident that ${\cal B}_q$ passes through the origin, 
$q=0$.  For real $a \ne 1$, ${\cal B}$ intersects the real $q$ axis at $q=0$
and at 
\beq
q_c(\{C\})=2(1-a)
\label{qcca}
\eeq
If, as is customary in physics, one 
restricts to $q \in {\mathbb Z}_+$, then the Potts antiferromagnet on 
$\{C\}$ has a zero-temperature phase 
transition if and only if $q=2$ (Ising case).  In the ferromagnetic case,
${\cal B}_q$ does not cross the positive real $q$ axis. 
For $q \ne 0,2$, the 
slice of this submanifold in the $u$ plane forms the circle 
\beq
{\cal B}_u: \ \ u=(q-2)^{-1}(-1+e^{i\omega}) \ , \quad \omega \in [0,2\pi) \ ,
\quad q \ne 0, 2
\label{ucircle}
\eeq
while for $q=2$, ${\cal B}_u$ is the imaginary $u$ axis.  For $q=0$, $Z=0$ and
no ${\cal B}_u$ is defined. 
We have obtained similar results for other graphs, including cyclic ladder
graphs $L_m$ \cite{a}. The corresponding Tutte polynomial 
obeys a recursion relation given in \cite{bds}.   Some zeros in the $q$ plane
are shown in Figs. \ref{ladqa0p25} and \ref{ladqa0p5}:

\newpage

\begin{figure}
\vspace{-4cm}
\centering
\leavevmode
\epsfxsize=4.0in
\begin{center}
\leavevmode
\epsffile{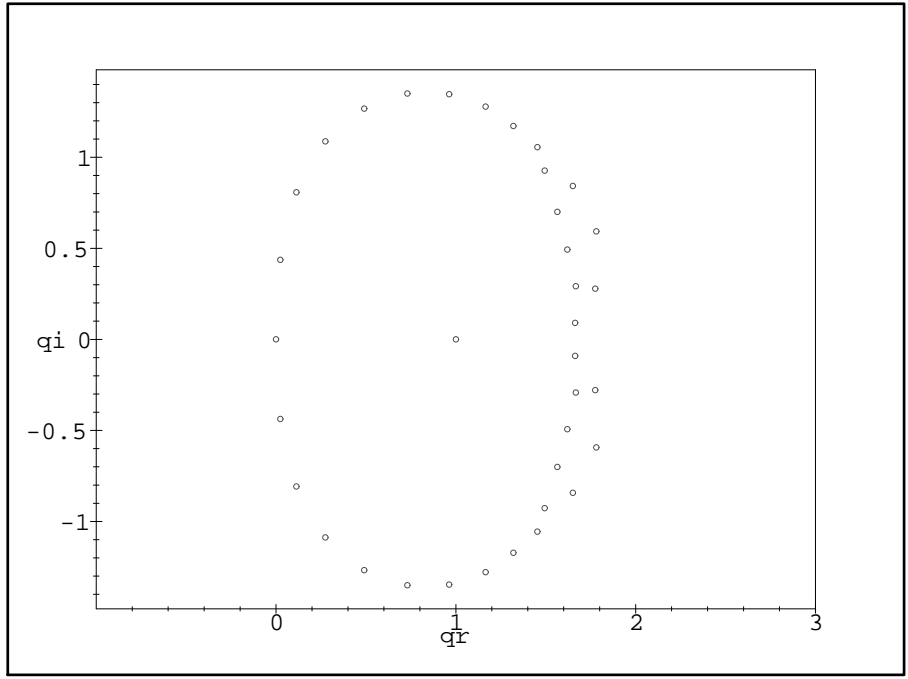}
\end{center}
\vspace{-2cm}
\caption{\footnotesize{Zeros of $Z(L_m,q,a)$ in $q$ for $m=18$ and
$a= 0.25$.  The axis labels $qr \equiv Re(q)$ and
$qi \equiv Im(q)$.}}
\label{ladqa0p25}
\end{figure}

\begin{figure}[hbtp]
\vspace{-4cm}
\centering
\leavevmode
\epsfxsize=4.0in
\begin{center}
\leavevmode
\epsffile{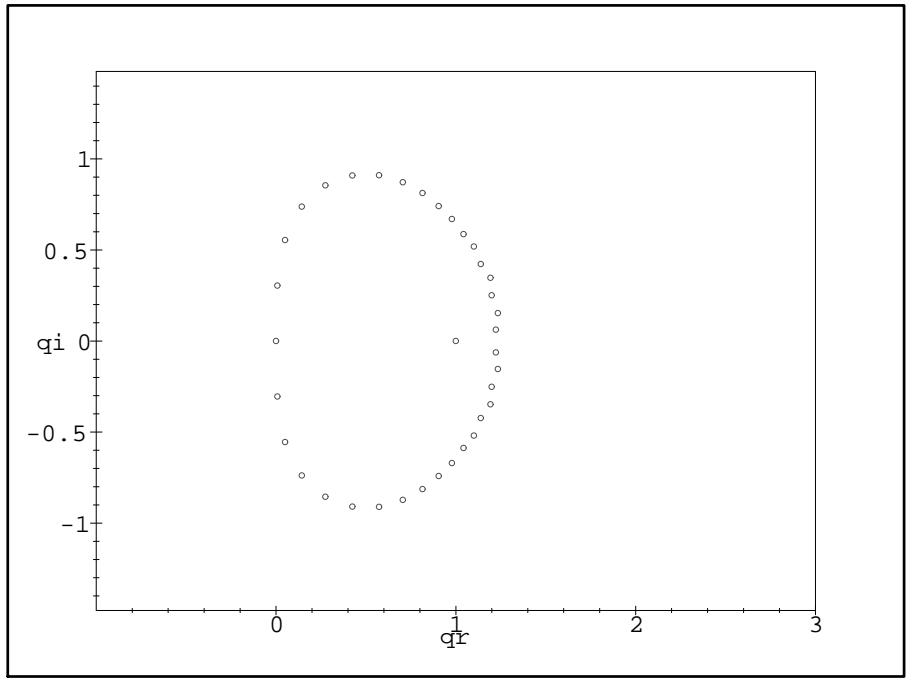}
\end{center}
\vspace{-2cm}
\caption{\footnotesize{Zeros of $Z(L_m,q,a)$ in $q$ for $m=18$ and
$a=0.5$.}}
\label{ladqa0p5}
\end{figure}

The locus ${\cal B}$ always passes through the origin, $q=0$.  
For $0 \le a < 1$, 
${\cal B}$ crosses the positive real $q$ axis at $q_c$, where
\beq
q_c(\{L\}) = (1-a)(a+2) \quad {\rm for} \quad 0 \le a \le 1
\label{qclad}
\eeq
and separates the $q$ plane into several regions. 
As $a$ increases from 0 to 1, the locus ${\cal B}$ 
contracts monotonically toward the origin, $q=0$ and in the limit as $a \to
1$, it degenerates to a point at $q=0$.  This also describes the general
behavior of the partition function (dichromatic) zeros themselves.
That is, for finite graphs, there are no isolated partition function zeros
whose moduli remains large as $a \to 1$. This is clear from continuity
arguments in this limit, given eq. (\ref{za1}).

\section{Rigorous Bounds on $W$}

Using a coloring matrix method, Biggs \cite{b77} proved upper and lower bounds
for $W$ for the square lattice.  We applied this method to prove such bounds
for the triangular and honeycomb lattices \cite{ww}.  These bounds are very
restrictive even for moderate $q$.  Although a bound on a given function need
not, {\it a priori}, coincide with a series expansion of that function, the
lower bound for the honeycomb lattice coincides with the first eleven terms of
the large--$q$ expansion for $W(hc,q)$.  We have proved a general lower bound,
which is applicable for any Archimedean lattice \cite{wn}. Here an Archimedean
lattice is defined as a uniform tiling of the plane with one or more regular
polygons such that all vertices are equivalent to each other.  It can be
specified by the ordered sequence of polygons $p_i$ traversed by a circuit
around any vertex: $\Lambda = \prod_i p_i^{a_i}$.  Let $\sum a_i = a_{i,s}$ and
$\nu_i= a_{i,s}/p_i$.  Then our general lower bound is \cite{wn}
\beq 
W \ge \frac{\prod_i D_{p_i}(q)^{\nu_{p_i}}}{q-1}
\label{wlbarch}
\eeq
For the function 
\beq
\overline W(\Lambda,y) = \frac{W(\Lambda,q)}{q(1-q^{-1})^{\Delta/2}}
\label{wbar}
\eeq
this bound reads
\beq
\overline W \ge 
\prod_i \Bigl [ 1+(-1)^{p_i}y^{p_i-1} \Bigl ]^{\nu_{p_i}}
\label{wbarlbarch}
\eeq

\vspace{4mm}

We are grateful to S.-H. Tsai for collaboration on many papers and to
M. Ro\v{c}ek and N. L. Biggs for the collaborations in \cite{strip} and
\cite{tk}. Our research was supported in part by the NSF grant PHY-97-22101.
Some further relevant work since BCC99 on wider strip graphs is in
\cite{ps,ss}.

\vfill
\eject

\end{document}